\documentclass[journal,10pt]{IEEEtran}

\ifCLASSINFOpdf
\else
   \usepackage[dvips]{graphicx}
\fi
\usepackage{url}
\usepackage{algorithmicx}
\usepackage{algpseudocode}
\usepackage{algorithm}
\algdef{SE}[DOWHILE]{Do}{doWhile}{\algorithmicdo}[1]{\algorithmicwhile\ #1}%
\usepackage{amsmath}
\usepackage{dsfont, cuted}
\usepackage{amsfonts}
\usepackage{amsthm}
\usepackage[T1]{fontenc}
\usepackage[utf8]{inputenc}

\usepackage{subfigure}
\usepackage{flushend}
\usepackage{breqn}
\usepackage{multirow}
\usepackage{multicol}
\usepackage{booktabs}
\usepackage{fullwidth}
\usepackage[noadjust]{cite}
\usepackage{setspace}
\usepackage{hyperref}
\usepackage{bmpsize}
\usepackage{flushend}
\usepackage{xcolor}
\usepackage{soul}

\usepackage{graphicx}

\usepackage{amssymb}

\usepackage{multirow,tabularx}
\newcolumntype{C}{>{\centering\arraybackslash}X}
\newcolumntype{L}{>{\raggedright \arraybackslash}X}
\newcolumntype{R}{>{\raggedleft \arraybackslash}X}

\begin{document}

\title{Current Advances in Computational Lung Ultrasound Imaging: A Review}

\author{Tianqi Yang, Oktay~Karakuş,~\IEEEmembership{Member,~IEEE,} Nantheera Anantrasirichai,~\IEEEmembership{Member,~IEEE,}  Alin~Achim,~\IEEEmembership{Senior Member,~IEEE}
        \thanks{Tianqi Yang, Nantheera Anantrasirichai and Alin Achim are with the Visual Information Laboratory, University of Bristol, Bristol BS1 5DD, U.K.}
        \thanks{Oktay Karakuş is in the School of Computer Science and Informatics at Cardiff University. Cathays, Cardiff, CF24 4AG}
        \thanks{e-mail: tianqi.yang@bristol.ac.uk; karakuso@cardiff.ac.uk; n.anantrasirichai@bristol.ac.uk; alin.achim@bristol.ac.uk}
}

\markboth{IEEE Transactions Ultrasonics, Ferroelectrics, and Frequency Control on XXX, Vol. xx, No. x, 2022}
{Shell \MakeLowercase{\textit{et al.}}: Bare Demo of IEEEtran.cls for IEEE Journals}
\maketitle

\begin{abstract}
In the field of biomedical imaging, ultrasonography has become common practice, and used as an important auxiliary diagnostic tool with unique advantages, such as being non-ionising and often portable. This article reviews the state of the art in medical ultrasound image processing and in particular its applications in the examination of the lungs. First, we briefly introduce the basis of lung ultrasound examination. We focus on (i) the characteristics of lung ultrasonography, and (ii) its ability to detect a variety of diseases through the identification of various artefacts exhibiting on lung ultrasound images. We group medical ultrasound image computing methods into two categories: (1) model-based methods, and (2) data-driven methods. We particularly discuss inverse problem-based methods exploited in ultrasound image despeckling, deconvolution, and line artefacts detection for the former, whilst we exemplify various works based on deep/machine learning, which exploit various network architectures through supervised, weakly supervised, and unsupervised learning for the data-driven approaches.
\end{abstract}


\IEEEpeerreviewmaketitle

\section{\label{sec:1} Introduction}
\IEEEPARstart{I}{n} recent years, with the development of ultrasound (US) along with the advancement of medical imaging technologies, the examination and treatment performance of US in clinical applications, especially in the diagnosis of lung disease, has gradually been recognized and accepted by clinicians. This advancement has not only changed disease treatment, prognosis and patient management, but also became a visual stethoscope and auxiliary diagnosis tool that radiologists can rely on. US examination has consequently become an important medical imaging method, which is an effective supplement to other medical imaging modalities such as chest X-ray, chest computed tomography (CT), bronchoscopy, and magnetic resonance imaging (MRI). 

Among the various clinical applications of US, lung US (LUS) has been increasingly used as a support tool in not just lung but even nephrological and rheumatological disease diagnoses, and in fluid status assessment \cite{allinovi2017lung}. It is regarded as a prospective routine practice for bedside patient assessment \mbox{\cite{touw2015lung,touw2019routine}}. On the one hand, this is because LUS is a fast, portable, radiation-free, bedside and non-invasive imaging technique. On the other hand, LUS also helps in assessing the fluid status of patients in intensive care as well as in deciding management strategies for a range of conditions. Indeed, respiratory disease is extremely common, amounting to more than 414.6 million cases around the world \cite{FIRS}. Severe lung problems cause the death of more than 100 thousand people in the UK every year \cite{BLungFoundation}. According to the British Lung Foundation, somebody dies due to lung disease in the UK every five minutes \cite{BLungFoundation}. Lung disease is in fact the third common cause of death in the UK after heart disease and cancer  \cite{BLungFoundation}. Following the start of the COVID-19 pandemic in 2020, analysis and diagnosis of lung disease has become even more crucial. Rising demand has been spurred for timely and accurate diagnosis, as well as patient monitoring \cite{LUfuture}.

Well known image quality issues affecting ultrasonography (speckle noise, low resolution), together with the peculiarities and added difficulties of LUS investigations (small acoustic window, energy dissipation in air) make the tools of computational imaging ever more important in reconstructing information of the highest quality in order to rip the benefits of LUS in a variety of clinical applications.
This paper aims to provide a comprehensive review of the state-of-the-art computational LUS imaging approaches. On the one hand, we will review model-based methods, which are reasonably well studied in the relevant literature. 
On the other hand, with the current advances in machine learning (ML) and artificial intelligence (AI), approaches that learn representative models and features from a (usually high ) number of training data will be paid attention in this review, especially deep learning (DL) methods, because they have brought remarkable improvements to LUS image analysis as well as to LUS image quality. Since the current state of LUS development has not yet reached maturity, this review aims to provide inspiration and references for the development of yet more accurate and efficient computational LUS imaging approaches.

The remainder of this contribution is organised as follows: We describe the basic features and clinical applications of LUS imagery in Section \ref{sec:2}. Section \ref{sec:3} presents processing methods for the reconstruction of high quality ultrasound images within the two sub-classes of \textit{model-based} and \textit{data-driven} methods. Furthermore, Section \ref{sec:4} reviews model-based as well as data-driven approaches specifically developed for LUS image processing. The discussions in Section \ref{sec:5} serve the purpose of revealing future research directions by analysing current challenges, and Section \ref{sec:6} gives a short conclusion of the reviewed work. 

\section{\label{sec:2} Lung Ultrasound and its Clinical Applications}
This section is intended to provide a brief overview of medical ultrasound imaging technology, with a specific focus on LUS. We describe the devices that clinicians normally use, the standards that should be followed, and the lung artefacts that diagnosis is based on.

\begin{figure}[t]
\centering
\subfigure[]{\includegraphics[width=.24\linewidth]{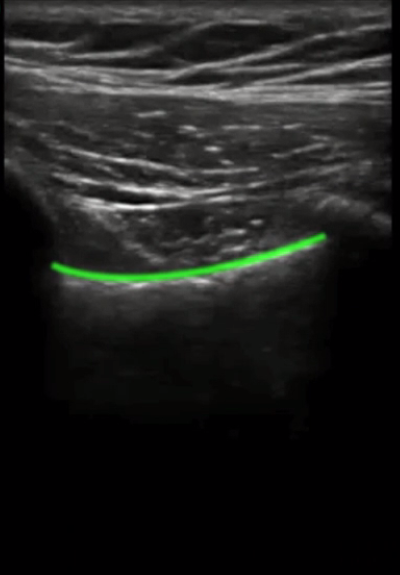}}
\subfigure[]{\includegraphics[width=.24\linewidth]{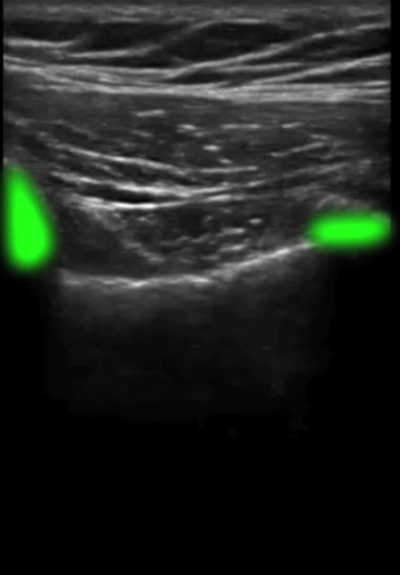}}
\subfigure[]{\includegraphics[width=.42\linewidth]{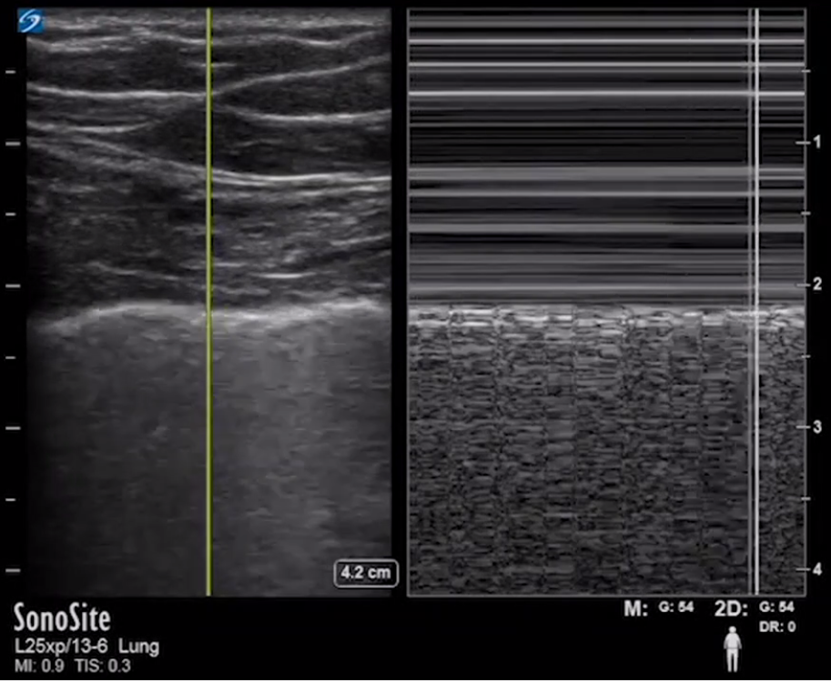}}
\hfill\\
\subfigure[]{\includegraphics[width=.30\linewidth]{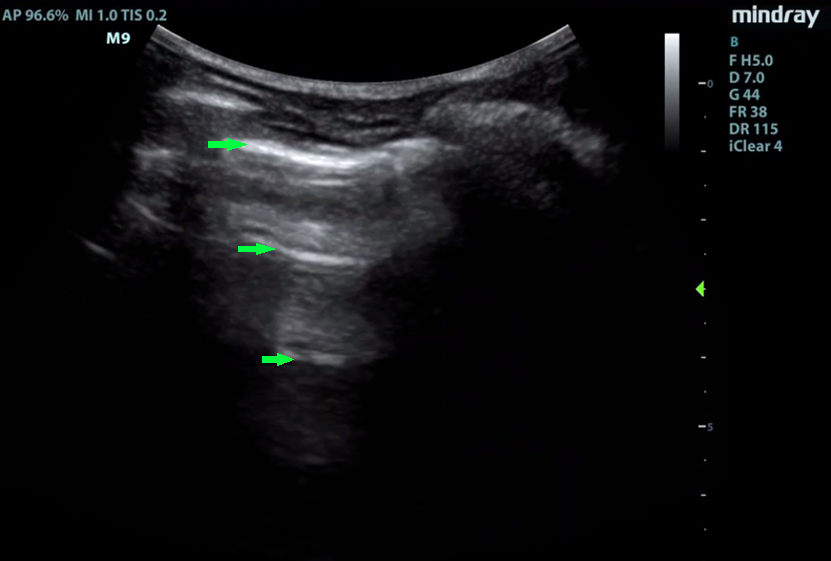}}
\subfigure[]{\includegraphics[width=.23\linewidth]{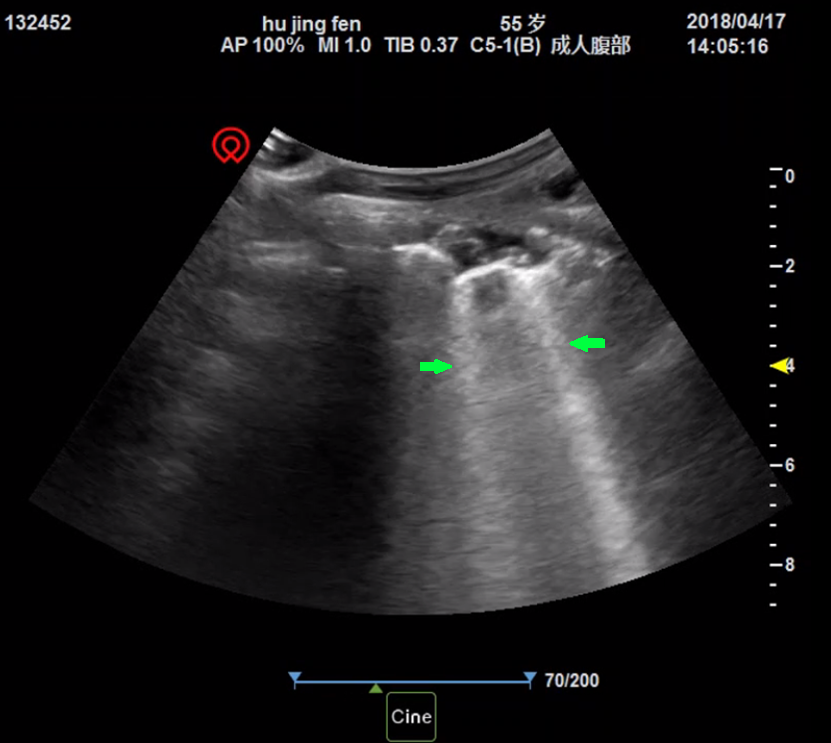}}
\subfigure[]{\includegraphics[width=.37\linewidth]{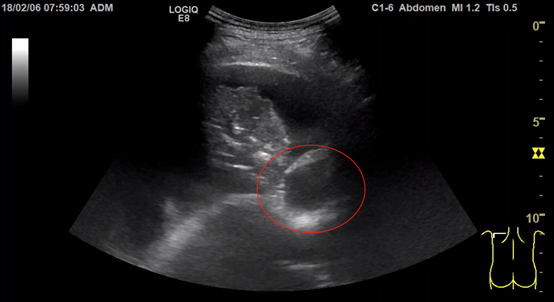}}
\hfill\\
\subfigure[]{\includegraphics[width=.29\linewidth]{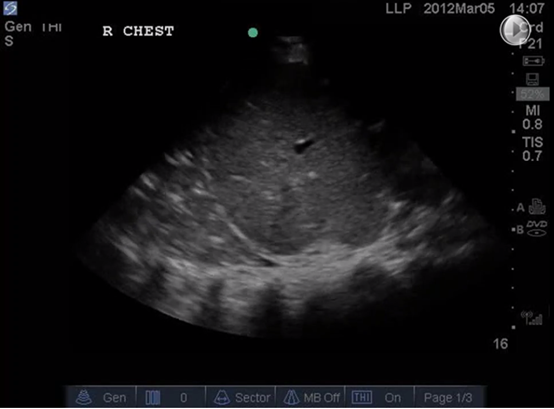}}
\subfigure[]{\includegraphics[width=.31\linewidth]{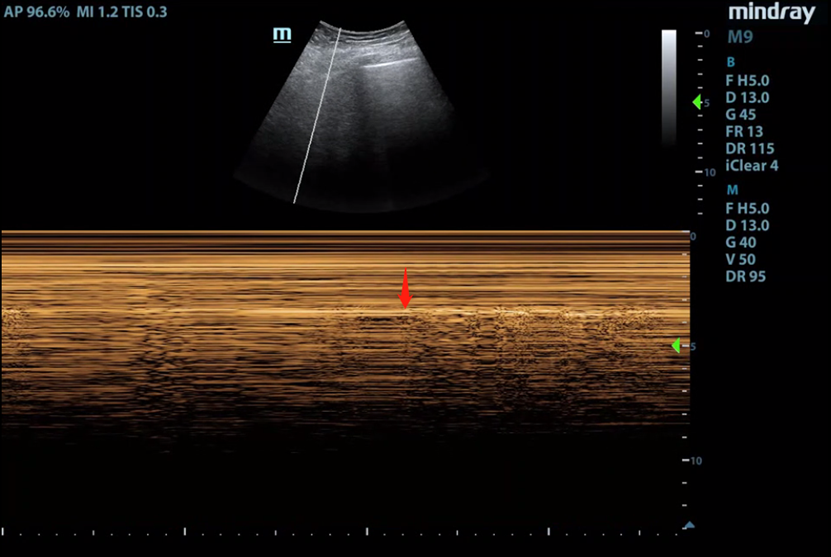}}
\subfigure[]{\includegraphics[width=.30\linewidth]{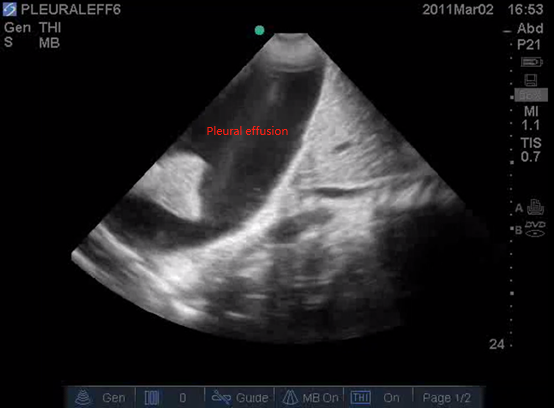}}\hfill\\

\centering
\subfigure[]{\includegraphics[width=.45\linewidth]{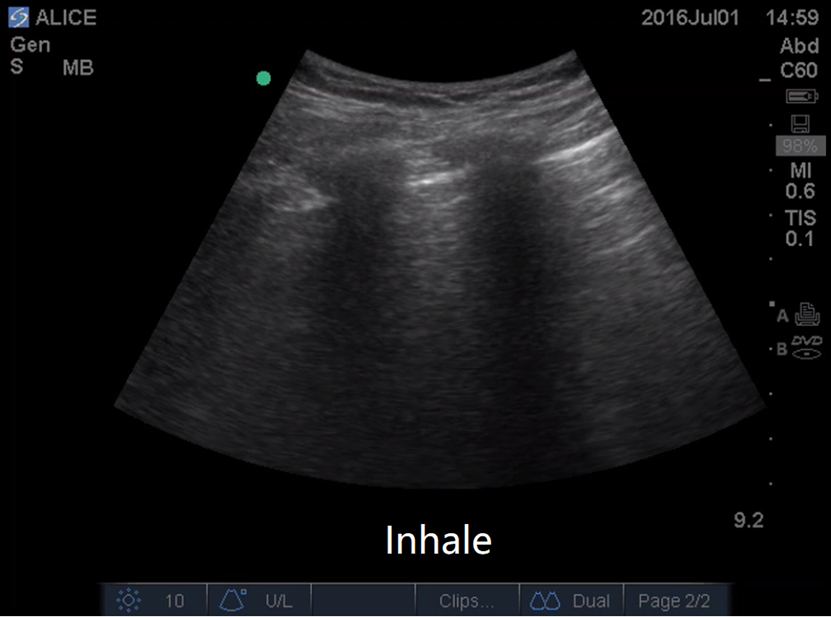}}
\subfigure[]{\includegraphics[width=.45\linewidth]{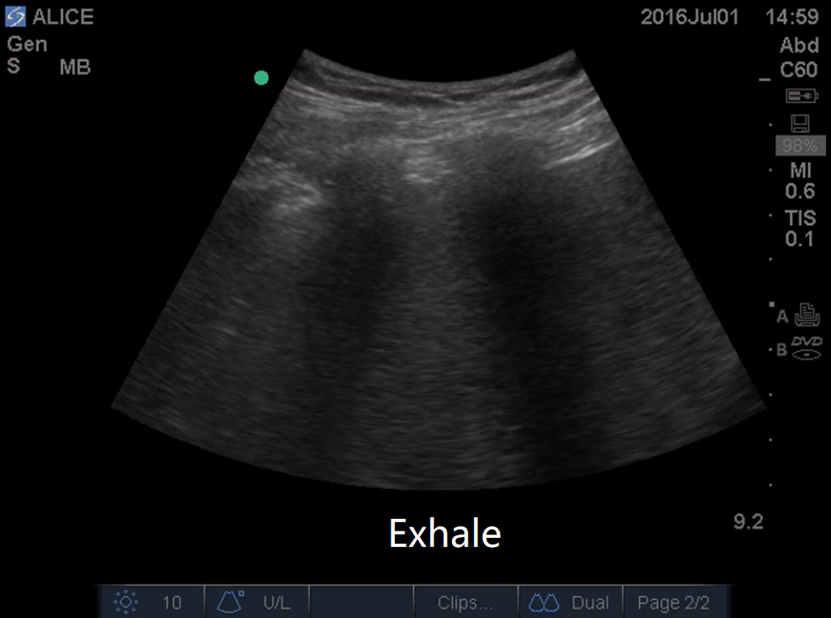}}

\caption{Example images for various types of Lung signs. \cite{Clinical} (a) Pleural line (linear probe), (b) Bat sign (linear probe), (c) Seashore sign (linear probe, B-mode on the left and M-mode on the right), (d) A-line (convex probe), (e) B-line (convex probe), (f) Atelectasis (linear probe), (g) Bronchial inflation (convex probe), (h) Lung point (convex probe, B-mode on the top and M-mode on the bottom), (i) Pleural effusion (convex probe),  (j) and (k) Curtain sign (convex probe).}
\label{signs}
\end{figure}

\subsection{\label{subsec:2:1} Instrumentation}

When sound waves are transferred into the body, they are transmitted until they are completely attenuated, the echoes received from the different depths can be envelop detected and displayed, and hence cross-sectional US images of the human body can be formed \mbox{\cite{2016Inverse}}.
 
 A widely accepted linear model of US image formation was proposed in \mbox{\cite{USimagingmodel}} based on the physical propagation characteristics of sound waves. The received signal can be expressed in the time domain as
\begin{equation}
r(\textbf{r}_0,t) = h(\mathbf{r},t) \circledast f_m(\mathbf{r})|_{\mathbf{r}=\mathbf{r}_0}\label{sound2image}
\end{equation}
where $\mathbf{r}_0$ represents the surface of the receiver, $\mathbf{r}$ refers to the location of the imaged tissue, $f_m(\mathbf{r})$ is the tissue reflectively function (TRF), and $h(\mathbf{r},t)$ refers to the system point spread function (PSF).

Probes are the actual ultrasonic transducers, which employ the piezoelectric effect to perform the bidirectional conversion of mechanical waves into electrical signals and vice versa. A non-exhaustive list of most commonly used clinical probes is given in Table \ref{tab:probes}.
In LUS imaging, for the observation of in-depth lesions, low-frequency convex array probes or low-frequency phased array probes are usually utilised, whereas high-frequency linear array probes are used for superficial pleural or sub-pleural detection. Image resolution increases with frequency, at the cost of penetration depth \cite{2019Application2}. When the density of the lung changes, artefacts which are used to distinguish sick and normal lungs can be detected, since some diseases can lead to an increase of the liquid content. Specifically, in a normal lung, the ultrasound beams disperse and become less organised as the waves travel within the alveoli that are full of air. This leads to fewer acoustic beams being reflected back to the probe, and therefore the region under the pleural line appears simply as a uniform background pattern finely sparkling with some linear echogenic horizontal lines \cite{volpicelli2013lung}. When the air content decreases and lung density increases due to the presence in the lung of transudate, the acoustic mismatch between the lung and the surrounding tissues is lowered, and the ultrasound beam can be partly reflected at deeper zones and repeatedly \cite{gargani2014lung}. This phenomenon creates discrete vertical hyperechoic reverberation artefacts. This provides clinicians the possibility to not only assess and evaluate conditions that are peculiar to the lungs but determine a fluid overload in the body. 

Depending on clinical applications, working principles, tasks and operating systems, LUS images can be acquired in different modes, mostly in M-mode and B-mode \cite{MedicalEquip}. An M-mode recording is conventionally displayed with the abscissa representing time, and the ordinate showing distance from the transducer, the latter derived from the time delay from echo emission to reflection and detection \mbox{\cite{carroll2021mmode}}. Since it can record rapid motions, it is usually used to detect the presence or absence of lung sliding. B-mode uses the amplitude of the echo pulse to modulate the brightness of the display. The abscissa and ordinate of the display correspond to the position of the sound velocity scan, thus forming an ultrasonic cross-sectional image modulated by the brightness \mbox{\cite{USbasis}}. It is a good choice for monitoring the existence of B-lines.

A number of vendors have developed LUS scanners, including the Canon Xario Series$\footnote{http://www.kangdamed.net/en/index.php/Xarioxilie/159.html}$, Aplio Series (by GE)$\footnote{http://www.kangdamed.net/en/index.php/Aplio/154.html}$ and Sonosite$\footnote{https://www.sonosite.com/products}$. In order to satisfy eventual point-of-care (PoC) needs, many of these vendors have been developing portable devices, such as Clarius and Butterfly iQ+, whose apps support high-resolution real-time lung monitoring under the choice of various modes.

\subsection{\label{subsec:2:2} LUS Standardization}

The imaging devices and the management of the patients varies in different countries. Hence, to promote an effective global medical exchange, a standardization for the international protocol of LUS is needed. Soldati et. al. \cite{soldati2020proposal} had developed a standardized approach for the use of equipment and the acquisition protocol, specifically for COVID-19 patients. A scoring system for severity classification was also proposed. An artificial intelligence system was developed to support the generation of standardized datasets \cite{LUfuture}.

Another standardisation effort was made by Allinovi and Hayes \cite{allinovi2021simplified} who demonstrated that a simplified 8-site B-lines score holds an almost identical predictive power to a 28-site score in quantifying the extent of extravascular lung water in children. The use of 8-site lung ultrasound assessments for children on dialysis was therefore proposed, to accurately quantify fluid overload in children.

\subsection{\label{subsec:2:3} Lung Signs and Conditions}

The lungs are the main air-containing organs of the human body. In a pathological condition, the air-liquid ratio in the lungs changes, and abnormal artifacts may appear during a US examination (see examples in Figures \ref{signs}). The analysis of LUS images is based on these signs. 

The common features in all clinical conditions, both local to the lungs (e.g. pneumonia, chronic obstructive pulmonary disease (COPD)) and those manifesting themselves in the lungs (e.g. kidney disease, COVID-19) are the presence in LUS of two types of artefacts, known as A- and B-lines. These may carry important information about the severity of diseases. Hence, the majority of research conducted in this area starts from detecting and quantifying these linear features in LUS images. The A-lines (horizontal linear structures) are a kind of artifact caused by multiple reflection of sound waves due to the difference of acoustic impedance between the pleura and the lung. Their presence are indicative of a healthy lung, whereas the B-lines, appearing as a vertical comet-tail artifact arising from the pleural line, relate to decreased lung aeration \mbox{\cite{soldati2015lung}} and can be indicative of multiple diseases \mbox{\cite{van2019b,Wang2021,allinovi2021simplified}}. Therefore, detecting B-lines has become common practice when evaluating various diseases with LUS  \cite{Ramin2018Automatic}. The position of an artifact relative to the pleural line (which is the horizontal echo reflection formed by the surface of visceral and parietal pleura) is also important in determining whether the detected artifact is a B-line or merely air or another organ \cite{2019DEVICES}. Therefore, in some studies, the pleural line is used as reference for the positioning of other line artefacts  \cite{2020Automatic,2018Automatic}. 

Indirect signs in LUS can reveal the severity of lung disease. In Table \ref{tab:lungconditions}, we present a non-exhaustive list of diseases, observable with LUS, along with their peculiarities. Apart from the listed conditions, LUS plays important role in intensive care units (ICU) conditions. It can effectively detect lung consolidations with a thickness greater than 20 mm in ICU patients \cite{Lichtenstein2016Lung}, evaluate the respiratory system and circulatory system in real time. LUS is suitable for detecting neonatal respiratory distress syndrome (NRDS), meconium aspiration syndrome, acute pneumothorax and occult atelectasis as well. It is also worth noting that LUS signs have also been shown to become a good indicator of extravascular lung water \cite{2019Quantifying}, and fluid overload in infants and children on dialysis when compared to other measures \cite{allinovi2016finding, allinovi2017lung}. 

\begin{table*}[ht]

\scriptsize
  \centering
  \vspace{0.3cm}\caption{Ultrasound probes\cite{rippey2020lung}}\setstretch{0.9}\scriptsize
    \begin{tabular}{@{}|p{4cm}|p{12cm}|@{}}  
    \toprule
    \textbf{Probe} & \textbf{Applications} \\
    \toprule
    The curvilinear transducer & This is the most commonly used probe for lung ultrasound, operating at relatively low frequency (1 \~{} 5MHz). It has good penetration and enables examination of a larger field of the pleural surface. \\     \hline
    
    The linear transducer &  This transducer operates at a higher frequency, it is good at interrogating a pleural surface with a maximal depth of 4cm or so. One selects this probe when wanting the highest resolution examination of an abnormality at the pleural surface. \\ \hline
    
    The phased array transducer & This is a relatively low frequency phased array transducer (1 \~{} 3MHz), which becomes a reasonable selection for lung ultrasound where gross and diffuse pathology is suspected. This would include pulmonary oedema and large pleural effusion or pneumothorax detection.  \\\hline

    \hline
    \end{tabular}%
    \label{tab:probes}%
\end{table*}%
 
\begin{table*}[ht]
\scriptsize
  \centering
  \caption{Lung signs}\setstretch{0.9}\scriptsize
    \begin{tabular}{@{}|p{5cm}|p{11cm}|@{}}  
    \toprule
    \textbf{Sign} & \textbf{Details} \\
    \toprule
    Pleural line and pleural sliding &  The horizontal linear high echo below the ribs constitutes the pleural line. The relative motion of the parietal pleura and the visceral pleura forms a reciprocating movement with breathing. \\ \hline
    
    A-line & In B-mode, A-lines are a series of hyperechoic lines that are equally spaced and parallel to the pleural line and occur below the pleural line. \\\hline
    
    Quadrilateral sign and sine wave sign & Below the pleural line, a line regular and roughly parallel to the pleural line (the lung line), together with the pleural line and the shadow of the ribs, display a quad sign. In M-mode, the movement of the lung line is sinusoidal \cite{Lichtenstein2016Lung}.\\\hline
    
    B-line & A vertical artifact with clear boundaries and moving synchronously with lung sliding. \\\hline
    
    Lung consolidation and atelectasis &  The echogenic US appearance of the lungs is similar to that of the liver or spleen. \\\hline
    
    Bronchial inflation sign &  A dividing point seen in real-time ultrasound where the lung sliding exists and disappears alternately with breathing movement. \\\hline
    
    Lung points &  A dividing point seen in real-time ultrasound where the lung sliding exists and disappears alternately with breathing movement. \\\hline
     
    Curtain sign & As air-filled lung moves up and down with the breathing motion and other abdominal organs are covered periodically. \\\hline

    \hline
    \end{tabular}%
    \label{tab:lungsigns}%
\end{table*}%

\begin{table*}[ht]
\scriptsize
  \centering
  \caption{Lung conditions}\setstretch{0.9}\scriptsize
    \begin{tabular}{@{}|p{4cm}|p{12cm}|@{}}  
    \toprule
    \textbf{Condition} & \textbf{Details} \\
    \toprule

    Pneumothorax &   The disappearance of the pulmonary sliding sign in the initial stage and the M-mode LUS shows a superposition of parallel lines that lack motion characteristics  \cite{Lichtenstein2016Lung}. \\ \hline
    
    Pneumonia and COVID-19 & Varying degrees of lung consolidation and abnormal pleural lines, some with fusion of B-lines and pleural effusion. Early stage of COVID-19 pneumonia shows single and/or confluent vertical artifacts \cite{LUfuture,2020What,2020Point}. Further evolution of COVID-19 pneumonia shows evident consolidations and widespread patchy artifactual changes \cite{soldati2020there}. Moreover, ultrasound elastography, a technique developed to measure elastic properties of superficial lung tissue, has shown ability to evaluate lung disease and potential for COVID-19 pneumonia diagnosis \mbox{\cite{zhou2020ultrasound}}. \\ \hline
    
    Chronic  obstructive  pulmonary  disease(COPD) & Pulmonary A-line, pulmonary sliding sign and no right ventricular overload \cite{Clinical}.\\ \hline
    
    Acute respiratory distress syndrome (ARDS) & Lung consolidation with bronchial inflation sign, abnormal pleural line, diffuse pulmonary edema, and disappearance of A-lines. extra-vascular lung water (EVLW) is another indicator to classify the severity of ARDS \cite{2018Wu}.\\ \hline
    
    Lung cancer &  US can clearly captures the chest wall, pleura and peripheral lung lesions, and can show the morphology, boundary and blood flow of the lesions and the anatomical relationship between fine structures and the surrounding tissue, which provides basic information for clinical diagnosis \cite{Clinical}. Through their US morphologies, one can distinguish benign and malignant lung tumors. \\ \hline
    
    Congenital diaphragmatic hernia (CDH) &  \textit{i.} partial absence of the hyperechoic line representing the normal diaphragmatic profile, \textit{ii.} partial absence of the pleural line in the affected hemithorax, \textit{iii.} absence of A lines in the affected area, \textit{iv.} presence of multi-layered area with hyperechoic contents in motion (normal gut), and \textit{v.} possible presence of parenchymatous organs inside the thorax (i.e., liver or spleen). \\ \hline

    \hline
    \end{tabular}%
    \label{tab:lungconditions}%
\end{table*}%

\section{\label{sec:3} LUS Image Enhancement and Reconstruction}
In order to accurately identify the various signs of disease in LUS images, the reconstruction of high quality images through observations and the extraction of information therein are necessary. Typically, this process involves solving an inverse problem whereby a set of unknown deterministic parameters which are observed through a linear transformation and corrupted by noise are estimated \cite{Eldar2005RMSE}. The solution however may not be unique, or it may be highly sensitive to changes in the data, which illustrates the ill-posedness of the problem. In the following, we discuss the most widely studied inverse problems in conjunction with medical US image analysis, which are also relevant to LUS. Approaches peculiar to LUS specifically are discussed in Section \ref{sec:4}.

\subsection{Model-Based Methods}\label{subsec:3:1}
\subsubsection{Despeckling}\label{subsubsec:3:1:1}

Generally, the final envelope detected LUS image is composed of two elements, the useful signal component (corresponding to structure inside the human body) and the noise component (comprising multiplicative speckle and additive measurement noise).
\begin{equation}
f(x,y) = g(x,y)n(x,y)+w(x,y),  \quad(x,y)\in Z^2
\label{despeckling-1}
\end{equation}
where $g(x, y)$ and $f(x, y)$ represent the speckle free and the observed signals, respectively. $n (x, y)$ and $w (x, y)$ represent the multiplicative and additive noise components, respectively. $(x, y)$ are the two-dimensional spatial coordinates. Since the influence of the additive noise is far less obvious than that of the multiplicative noise \cite{Achim2001Novel}, the image formation model \eqref{despeckling-1} can usually be approximated by
\begin{equation}
f(x,y) = g(x,y)n(x,y),  \quad \text{where }(x,y)\in Z^2. \label{despeckling-2}
\end{equation}
The term $n(x,y)$ corresponds to speckle noise, which is an inherent phenomenon in LUS images. The early work described speckle noise characteristics in the wavelet domain with a heavy-tailed $\alpha$-stable distributions \cite{Achim2001Novel}. 

Speckle noise has been shown to be correlated with the tissue structure \cite{Achim2001Novel}, so its statistical description generally depends on the type of tissue and the imaging system. It exhibits granular patterns, which obscure fine anatomical details, and thereby reduce the diagnostic accuracy. Speckle noise is hence normally regarded as an undesirable phenomenon in most clinical applications. However, sometimes speckle noise can also constitute useful information, such as when used for speckle tracking (i.e. motion estimation) and lung tissue characterization \mbox{\cite{2016Inverse,mohanty2020vivo,lye2021vivo}}. To mitigate speckle noise, a logarithmic transformation is usually employed to convert the multiplicative characteristics into an additive model.


Recently, Choi and Jeong \cite{2020Despeckling} have improved descpeckling performance by combining various approaches, including speckle reduction anisotropic diffusion (SRAD) filter\cite{Yong2002SRAD}, discrete wavelet transform (DWT) using symmetric characteristics, gradient domain guided image filtering (GDGIF) and weighted guided image filtering (WGIF). Compared with previous denoising methods, their algorithm showed superior despeckling performance, better feature information conservation, and lower computational cost. Another stat-of-the-art technique proposed by Chen et. al. \cite{2020chen} introduced the alternating direction multiplier (ADMM) algorithm \cite{Boyd2011Alternating} to optimize the denoising effect of the new speckle noise recovery model based on adaptive variational method. This computational framework for solving optimization problems is shown to be suitable for solving distributed convex optimization problems, and therefore is suitable for LUS image denoising tasks.

\subsubsection{Deconvolution} \label{subsubsec:3:1:2}
Another important post-acquisition operation often needed in ultrasonography is image deconvolution, whereby US images are modelled as a convolution between a blurring kernel or point spread function (PSF), and the tissue reflectivity function \cite{jensen1993deconvolution, Hourani2020,Hourani2021}. The linear image formation model can be rewritten as
\begin{equation}
y(r) = h(r)\circledast x(r)+n(r),  \quad r\in R,
\label{third}
\end{equation}
where $y(r)$ is the image pixel observed at position $r$, $x(r)$ is the TRF to be estimated, $h(r)$ is the system PSF, $n(r)$ is the additive measurement noise. $R$ refers to the image domain  \cite{2016Inverse}.

Deconvolution in LUS imaging is a tool to improve visual quality and achieve better contrast. This translates in easier interpretation for the physicians.  The most common strategies for medical US image restoration is represented by MAP-based deconvolution. It tackles the problem using a two-step scheme: the PSF is estimated first and subsequently image restoration is performed. The main advantages of these strategies lie in (i) two dimensional or even three dimensional PSFs can be accounted for \cite{taxt1995restoration, michailovich2005novel}, (ii) no assumption is made on the PSF, neither on the number of zeros or poles, and nor on the position of the zeros in the complex plane, and (iii) more advanced models than white Gaussian can be assumed for the tissue reflectivity  \cite{2011Statistical,besson2019physical}. Although commonly employed, MAP deconvolution brings an increased computational cost, even if simple schemes are employed such as based on Wiener filtering or $\ell_1$-norm optimization. Another problem is that the PSF estimation is still tedious. In particular this is due to the need for phase unwrapping procedures where non-minimum phase PSFs are considered.

Due to the inherent bandwidth limitations of US scanners and the adverse effects of measurement noise,  LUS image deconvolution is very sensitive to errors occurring in the PSF estimation. Even slight errors in the PSF estimates can lead to obvious artefacts that render the reconstructed images worthless. The method of Michailovich et. al. \cite{michailovich2019iterative} might be a possible solution to address this issue. They proposed a ``hybrid'' deconvolution technique, which was based only on partial information about the PSF, especially its power spectrum to estimate tissue reflectivity. While directly estimating the reflectance of the tissue from the relevant radio-frequency (RF) data, the proposed approach simultaneously eliminated errors caused by inaccuracies in PSF estimation.

The method proposed by Pham et. al. \cite{2020Joint} even overcame the main limitation of the former related to the requirement for PSF estimation.  This algorithm for dealing with a sequence of fast-changing US images was based on the combination of two different techniques: deconvolution robust principal component analysis (DRPCA) and blind deconvolution (BD). It provided similar performances as previous work, but the PSF was assumed to be spatial-temporally invariant, and the algorithm appeared to be computational complex.

\subsection{Data-Driven Approaches}\label{subsec:3:2}

\subsubsection{Despeckling} \label{subsubsec:3:2:1}
Deep learning methods are able to play a compelling role in improving the quality of US images, in terms of speckle mitigation in particular. Such data-driven systems can indeed be leveraged across the US imaging domain \cite{8808885}, including LUS.

The work in \cite{vedula2017ctquality} proved the applicability of CNNs as a method that can quickly and accurately perform image restoration. A multi-resolution fully convolutional neural network (FCN) was used to approximate an ultrasound image of ``CT quality''. This end-to-end framework effectively improved the image resolution and contrast while preserving all the relevant anatomical and pathological information. It offered low complexity, making it applicable in real-time settings. However, in practice, CT-US paired data is difficult to obtain, so the US data used for training were simulated. In real applications, the shortage of such multimodal data pairs can reduce the applicability of the method.
 
The method proposed by Feng et. al. \cite{BookSpeckle} was shown to retain all relevant anatomical and pathological information in the restored images. They introduced US-Net by enhancing the CNN structure detail and proposed a new hybrid loss function designed for speckle noise removal.


\subsubsection{Deconvolution} \label{subsubsec:3:2:2}
Deep neural networks have also found applications in ultrasound image restoration. Perdios et. al. \mbox{\cite{2017ADL2US}} made it possible to apply stacked denoising autoencoders (SDA) to the recovery of LUS images. They explored both a linear measurement case where a known Gaussian random matrix was used as the measurement matrix (SDA–CNL) and a non-linear measurement case where the weight matrices and bias vectors were learned (SDA–CL). It was shown that a 4-layer SDA–CL outperforms a state-of-the-art compressed sensing algorithm without the need to tune any hyper-parameter. While increasing the quality of the reconstructed image is the main objective, reducing the calculation time is also an important requirement in applications. Yoon and Ye \cite{2017DL4AccelUS} proposed a novel DL approach that interpolates the missing RF data by utilizing the sparsity of the RF data in the Fourier domain. The algorithm effectively reduced the data rate without sacrificing the image quality and is also applicable to LUS images. The CNN can be trained using a RF data measured by the linear array transducer for a particular organ, and can be extended for other types of transducers and/or different body parts.

\subsection{Hybrid Techniques}\label{subsec:3:3}
The key to image reconstruction is to find a good sparse representation of the image to be reconstructed (which is also true for model-based approaches). DL methods assume that the representation is provided by deep neural networks, so it is nonlinear. The model parameters of the representation must be learned from a large amount of data. Hybrid approaches were therefore proposed. In hybrid approaches, neural networks are normally used to learn the prior of the observed data such that the size of the required dataset can be reduced.

There are obvious theoretical connections between DL networks and traditional iterative algorithms. Jin et. al. \cite{2016Deep} explored the relationship between CNNs and iterative optimization methods for inverse problems where the normal operator associated with the forward model is a convolution. The proposed method, which is called FBPConvNet, combined filtered back projection (FBP) with a multiresolution CNN. The direct inversion realized by FBP encapsulated the physical model of the system, but led to artifacts when the problem is ill-posed; the CNN combined multiresolution decomposition and residual learning in order to learn to remove these artifacts while preserving image structure. The structure of the CNN was based on U-Net \cite{ronneberger2015u}, with the addition of residual learning. This approach was motivated by the convolutional structure of inverse problems in  biomedical imaging, including LUS.

Chang et. al. \cite{2017One} proposed a general framework to train a single deep neural network that solves arbitrary linear inverse problems. They observed that in optimization algorithms for linear inverse problems, signal priors usually appear in the form of proximal operators. Thus, the proposed network acted as a proximal operator for an optimization algorithm and projected similar image signals onto the set of natural images defined by the decision boundary of a classifier. The learned projection operator combined the high flexibility of deep neural nets with the wide applicability of traditional signal priors. This has the potential to lower significantly the costs involved in the design of specialized hardware, where LUS is a clear potential beneficiary.

Similar to other US images, LUS images exhibit speckle and require pre-processing. With virtually no modifications, the aforementioned despeckling and deconvolution techniques can be adopted specifically for LUS imaging, and therefore contribute to better feature extraction in high resolution LUS images.

\section{\label{sec:4}Computational LUS Image Analysis}

This section presents the latest advances in computational imaging and image analysis methods utilising LUS. The LUS image processing methods are grouped into two major categories, specifically into the long developed \textbf{model-based} approaches and the \textbf{data-driven} (or learning-based) methods. In subsection \ref{subsec:4:1}, we talk about LUS line artefacts identification in a model-based way. Subsection \ref{subsec:4:2} reviews data-driven approaches that have been developed in clinical LUS interpretation including artefact localization, segmentation and so on. 

\subsection{\label{subsec:4:1} Model-based Methods}

Most tasks of LUS analysis are related to the identification of line artefacts. An added benefit of model-based approaches is that they constitute an unsupervised framework for the labelling of LUS images (e.g. B-lines) in cases where annotated data are not available for machine learning. 

To mask line artefacts on LUS images, the authors in \cite{Ramin2018Automatic} generated a normalized gray scale map that can be used for delineation of different structures in the images. They used a random walk method to delineate the pleural line, and then excluded the upper pleural region before identifying B-lines. This was achieved by an alternate sequential filtration, and subsequently applying the top-hat filter to ensure that B-lines are laterally detached. Finally, a Gaussian model was fitted to each detected B-line, and the peak point of the fitted Gaussian models corresponding to the B-lines are calculated and used to accurately determine the position of B-lines. B-lines were then overlaid on the B-mode images. Similarly, to implemented automatic pleural line detection, the authors of \cite{2020Automatic} masked a LUS image with expectation maximization (EM) based thresholding, after which hidden Markov model (HMM) and Viterbi algorithm (VA) were used to highlight the pleural line. 

Line detection in LUS can also be posed as an inverse problem, whereby the line information (position and orientation) is estimated from the observations. The solution however may not be unique, or it may be highly sensitive to variations in the data, which illustrates the ill-posedness of the problem. Regularization with specific penalty functions (or by employing specific priors if the problem is addressed in a Bayesian framework), is then necessary to solve such problems. Normally, the regularization term is related to the prior information of the parameter to be estimated \cite{2016Inverse}. In the literature, there are three main categories of such approaches, which include (i) statistical methods, (ii) regularized geometric modeling methods, and (iii) methods based on sparse representations \cite{illpose} or, sometimes, a combination thereof. In the following, we discuss the most widely studied inverse problems in conjunction with LUS image analysis. 

In the very first work adopting an inverse problem formulation, Anantrasirichai et. al. \cite{2016Line} proposed an innovative way of detecting line artefacts in LUS images by employing the Radon transform, converting an image into a space of lines, and solving the following optimization problem:
\begin{align}
    \hat{x} = \arg\min_{\hat{x}} \left\{\frac{1}{2}\|y - \mathcal{R}^{-1}x \|_2^2 + \alpha\|x\| + \beta\|\bigtriangledown\mathcal{R}^{-1}x \|\right\},
\end{align}
where $\mathcal{R}$ represents Radon transform, $\alpha$ and $\beta$ are regularisation constants and $\hat{x}$ is the set of detected lines. The ADMM algorithm was used to solve this optimisation problem, offering a fast convergence rate. The scheme firstly detected the pleural line in order to locate the lung space. Then, the local peaks of the Radon transform were detected and line-type classification was done following clinical definitions, in the spatial image domain. B-lines, A-lines and Z-lines were hence successfully identified. In \cite{2017Line}, the authors further extended the method by combining the Radon transform with the PSF of the US acquisition system in a single equation thereby achieving line detection and deconvolution simultaneously. To enhance line detection performance and the visualization of restored lines, they included an additional convolution factor in the Radon transform domain with an unknown blurring kernel. The penalty function employed was the $\ell_p$-norm with norm order values of $(0 \le p \leq 1)$ in order to promote sparsity in the Radon space. 

Extending the above work in the context of evaluating COVID-19 patients, Karakus et. al. \cite{2020Detection} improved the line detection performance by regularizing the solution using the Cauchy proximal splitting (CPS) algorithm \cite{Karakus2020} to promote statistical sparsity by utilising the Cauchy-based penalty function.

\begin{figure}[ht]
\centering
\subfigure[]{\includegraphics[width=.45\linewidth]{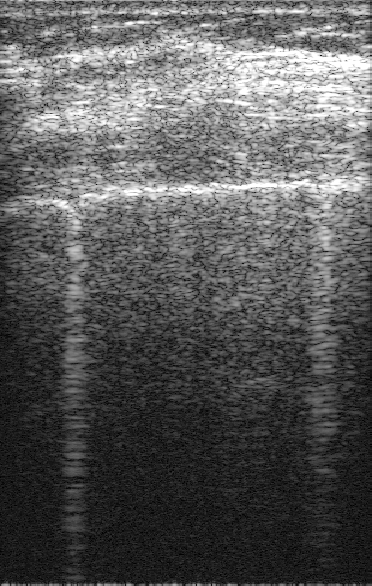}}
\subfigure[]{\includegraphics[width=.45\linewidth]{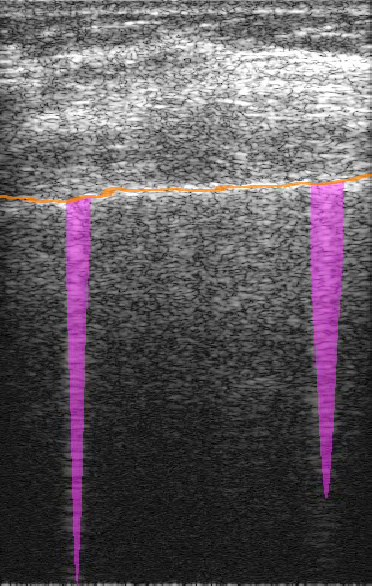}}
\caption{B-line detection results of Moshavegh et. al. \cite{Ramin2018Automatic}. (a) LUS scan containing two B-lines. (b) Pleural line is outlined and two B-lines are detected in (a) and overlaid.}
\label{Ramin_automatic}
\end{figure}

\begin{figure}[ht]
\centering
\includegraphics[width=.9\linewidth]{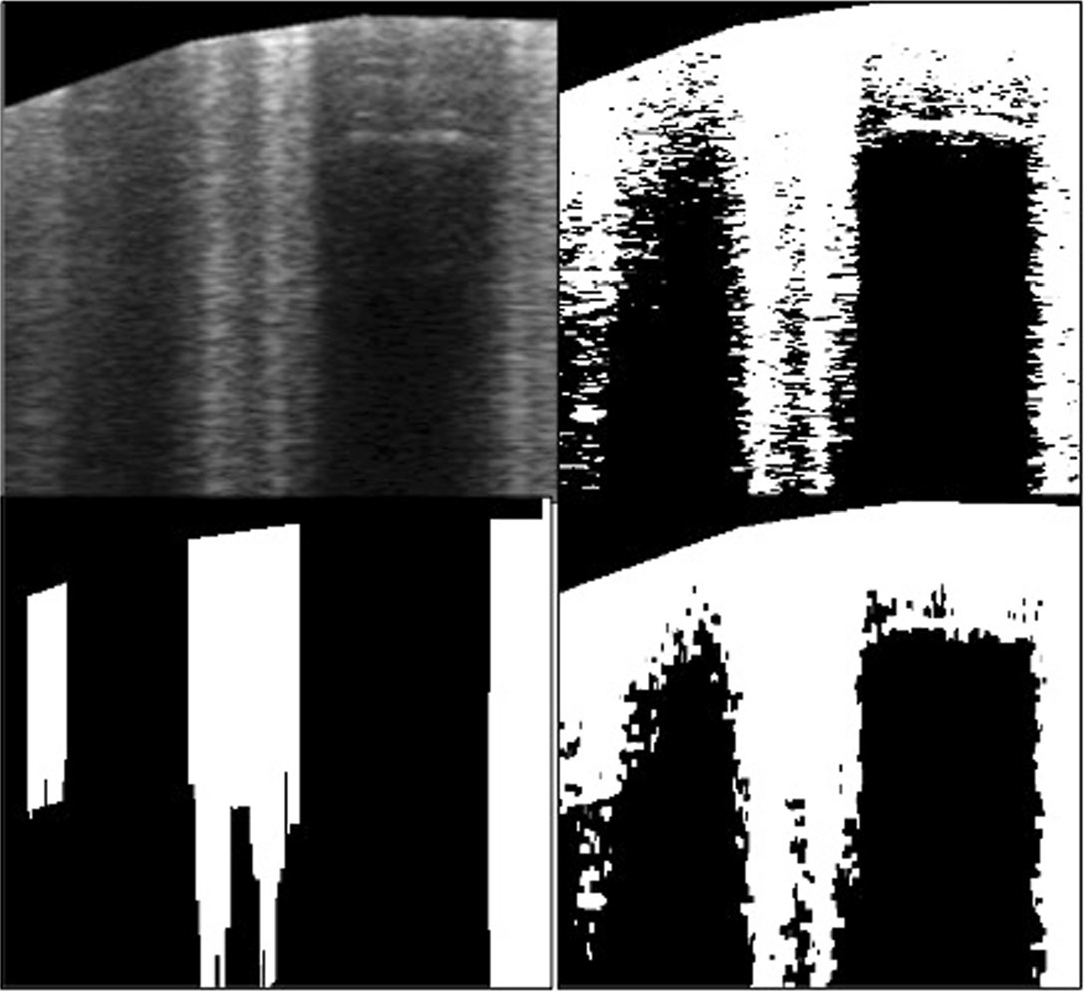}
\caption{Segmentation algorithm steps of Brusasco et. al.  \cite{brusasco2019quantitative}. Upper left: sub-pleural selection from the original ultrasound scan. Upper right: K-means classification to divide pixels into two subsets. Bottom left: B-line detection. Bottom right: alternated sequential filter consisting in iterative morphological openings and closings.}
\label{Brusasco_quantitative}
\end{figure}

\begin{figure}[ht]
\centering
\includegraphics[width=.7\linewidth]{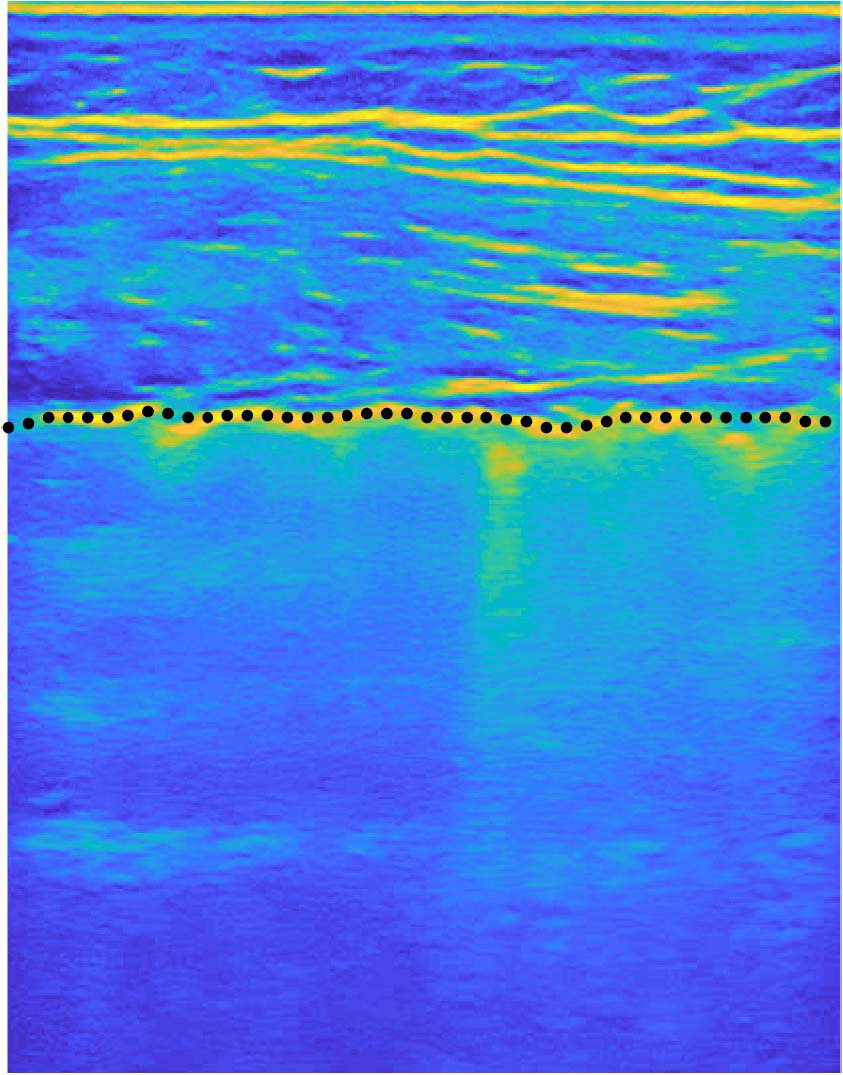}
\caption{Pleural line (black dotted line) detection result of Leonardo et. al.\cite{2020Automatic}.}
\label{carrer_automatic}
\end{figure}

\begin{figure}[ht]
\centering
\includegraphics[width=.9\linewidth]{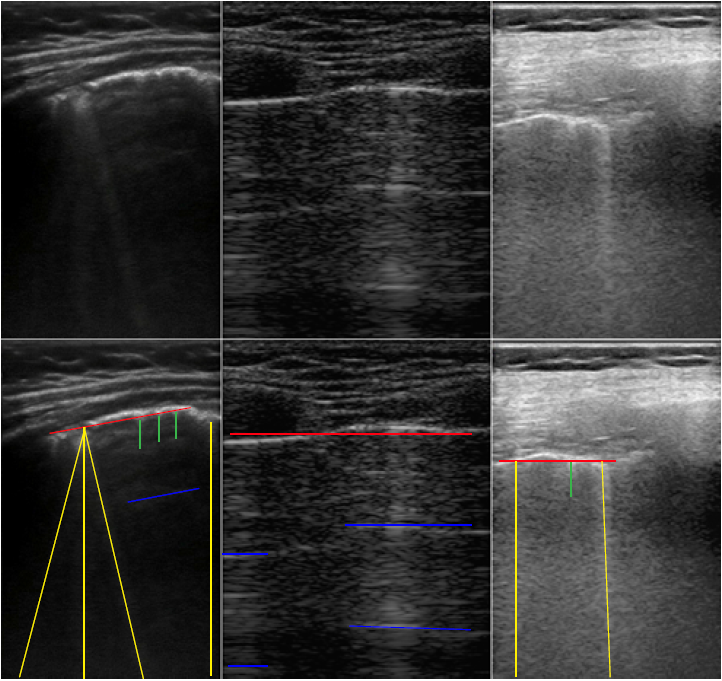}
\caption{Line artefacts in lung ultrasound images Red, yellow, blue and green lines represent the pleural lines, B-lines, A-lines and Z-lines,
respectively. \cite{2017Line}.}
\label{pui_line}
\end{figure}

\begin{figure}[ht]
\centering
\includegraphics[width=.9\linewidth]{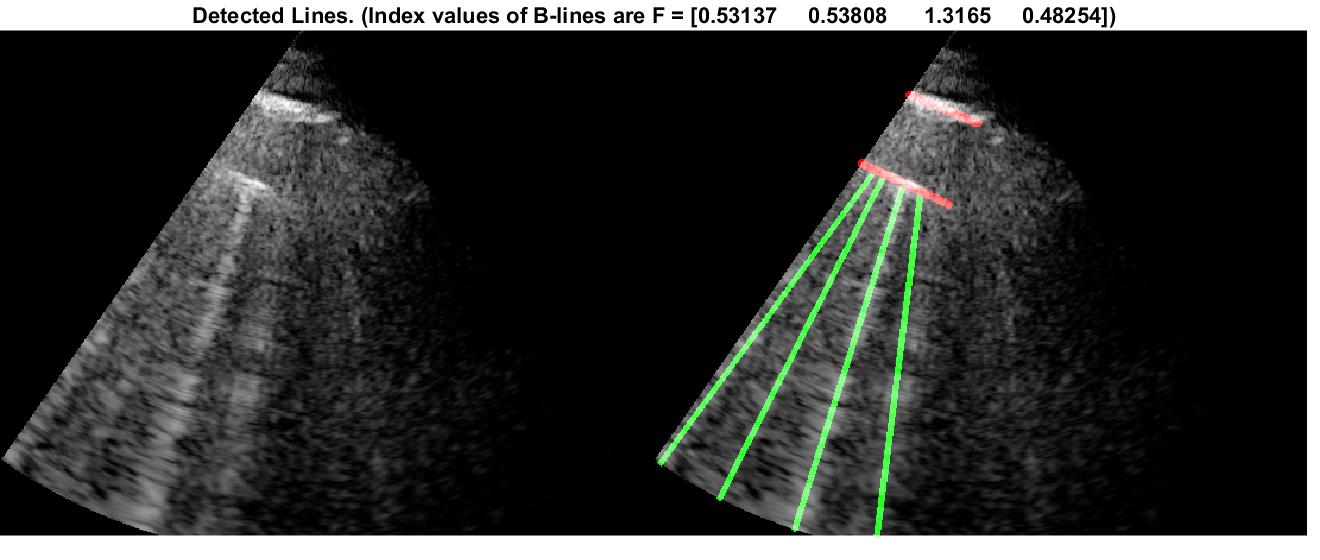}
\caption{Line detection results of Karakus et. al. \cite{2020Detection}. Left: original B-mode ultrasound image. Right: detected lines.}
\label{oktay_detection}
\end{figure}

\subsection{Data-Driven Approaches}\label{subsec:4:2}
Compared to conventional model-based approaches, ML based methods can provide improved performance in medical image analysis tasks, since they are able to capture more complex patterns in the data. ML approaches are often driven by patterns contained in the data, which are generally required for training and testing purposes. Therefore, the dataset used for ML is of great importance. Moreover, in a study on automated segmentation and scoring of COVID-19 LUS images, Roshankhah et. al. \mbox{\cite{roshankhah2021investigating}} also pointed out that the splitting strategy of the training and test datasets can influence the performance of the CNN.

In the sequel, we summarise the data-driven methods and split them into three main classes, fully-supervised, weakly-supervised and transfer learning.

\subsubsection{Fully-supervised Learning}\label{subsec:4:2:1}
Various studies in the literature have shown that the data-driven algorithms are effective tools for medical ultrasound image analysis, and they have already started to make an impact in LUS image analysis. As a first example, the work in \cite{2018Automatic} implemented a fully-supervised method, where a total of 1450 pneumonia and 1605 normal lung images were analyzed with a three-layer feed-forward neural network. The method combined image processing and vector classifiers to identify consolidations of pediatric pneumonia. After the identification of the pleural line and the removal of skin/soft tissue, feature extraction was carried out based on the analysis of brightness distribution patterns presented in rectangular segments from the images. In the end, the method was able to correctly identify pneumonia infiltrates with a sensitivity of 90.9\% and a specificity of 100\%. 

Later, Wang et. al.\cite{2019Quantifying} used a generic CNN structure to perform real-time lung B-line quantification to diagnose and quantify pulmonary edema. The training was implemented on a small dataset composed of 4864 clinical LUS images. The approach used (i) data augmentation to increase the variety of the data, (ii) dropout to avoid over-fitting, and (iii) rectified linear units (ReLU) to realize a faster training speed. Despite an apparently low 43.4\%  absolute accuracy in the test set, the intra-class correlation of 0.791 indicates substantial agreement of the neural network algorithm with the human-identified B-line counts. The B-line identification and clinical classification time was around 0.1s, nearly reaching real-time processing. 

Baloescu et. al. \cite{Cristiana2020Automated} proposed to use a relatively shallow custom-made network (CNN) architecture with 3-D filters (called 3D-CsNet) for the purpose of automatically distinguishing between the presence and absence of B-lines as well as assessing B-line severity. The backbone of 3D-CsNet is a CNN consisting of eight intermediate layers followed by two fully connected layer. As it alleviates the need for pre-training, the proposed algorithm is fast, employing few trainable parameters, and it is flexible and easy to deploy. However, this method has a weaker performance on multiclass severity rating (0.65 kappa) compared to its binary rating (0.88 kappa), because a higher number of categories may result in lower agreement when it comes to DL performance for the same amount of data.

To improve the diagnosis efficiency, the approach introduced by Kulhare et. al. in \cite{kulhare2018ultrasound} automatically segmented LUS features in simulated animal models with six single-class single shot detection (SSD) neural nets, making it suitable for on-device inference tasks. However, the success of application to human LUS images has not been investigated. 

Motivated by the COVID-19 pandemic, researchers have also been enthusiastically investigating various solutions to the application of the COVID diagnosis. McDermott et. al.  \cite{McDermott2021Sonographic} and Soldati et. al. \cite{soldati2020there} have illustrated the applicability of LUS imaging to COVID-19 symptoms detection. By comparing LUS protocols and image features, the former work suggested the possibility of aiding in interpreting LUS images autonomously or semi-autonomously, and the latter clarified the urgent need of the diagnostic and prognostic role of LUS in COVID‐19.
One of the supervised method examples is the quantitative and automatic LUS scoring system developed by Chen et. al. \cite{chen2021quantitative}, using multi-layer fully connected neural networks (FNNs) for evaluating COVID-19 pneumonia. The authors introduced a curve-to-linear conversion process, thereby reducing the feature extraction problem from two dimensions to one dimension. One and two-layer FNNs with hidden nodes were used to learn the data features. ADAM optimizer was applied to fit the data with ReLu as the activation function. With all the settings, the neural network with 128$\times$256 neurons gave the highest accuracy of 87\%.

\subsubsection{Weakly-supervised Learning}\label{subsec:4:2:2}
Even though the aforementioned approaches have brought improvements in LUS image interpretation, it cannot be denied that they require a high amount of labelled LUS images for training. This limitation has motivated researchers to develop weakly-supervised learning algorithms, where smaller amount of training samples are required. Van Sloun and Demi \cite{2019Localizing} explored weakly-supervised DL techniques, requiring only a single label per frame for training, and exploited gradient-weighted class activation mapping (grad-CAM) \cite{selvaraju2017grad} to perform B-line localization. The network has been applied and tested both in-vitro and in-vivo. Notably, the proposed method can also serve as a region-of-interest selector for further quantitative analysis of segmentation in the LUS data that contain the most relevant information.

Different from previous work, which relies on attention and CAMs to find the region in the image where the lesion is, the method proposed by Kerdegari et. al. \mbox{\cite{kerdegari2021b}} can localize B-line regions in LUS videos both spatially and temporally. They trained a RNN in a weakly-supervised manner. They leveraged temporal analysis networks, and used spatiotemporal attention to find the most important frames (i.e., B-line frames) and localized B-line regions within a video. Their proposed attention model achieved the highest F1 score of 83.2\% when benchmarked against previous techniques.

Subhankar et. al. \cite{2020DL4COVIDinPOC} presented an extended and fully-annotated version of the ICLUS-DB database\footnote{https://iclus-web.bluetensor.ai.}. They presented a deep network, derived from spatial transformer networks (STN) \cite{jaderberg2016spatial}. Two transformed versions of the input image are introduced to localize pathological artifacts. Then a feature extractor generates the final prediction. The network simultaneously predicts the disease severity score associated to an input frame and provides localization of pathological artefacts in a weakly-supervised way by inferring it from simple frame-based classification labels. In \cite{2020DL4COVIDinPOC}, a method based on uninorms was used to effectively aggregate the frame score at video level. An  F1-score prediction of 65.1 an was reported when performing the prediction at frame level.  For video-based score prediction, because of the low inter-annotator agreement and the small number of video-level samples annotation, an F1-score of 61\%, a precision of 70\% and a recall of 60\% were reported. When evaluating segmentation results, the pixel-wise accuracy reaches 96\% and a binary Dice score is 0.75. 
Tsai et. al. \cite{tsai2021automatic} further extended the idea of STN to Regularised Spatial Transformer Network (Reg-STN) via implementing a video-based (weakly supervised) and a frame-based (supervised) labelling approach. The video-based labelling approach reaches 91.12\% accuracy for 10-fold cross validation, and the frame-based reaches 92.38\%. This was the first attempt at full automation of LUS evaluation of lung pathology, and thus proposed an evaluation tool specifically for COVID-19. 

Inspired by the idea of support vector machines (SVM), which usually does not require a large number of training data, Veeramani and Muthusamy \cite{2015Detection} proposed a method called relevance vector machines (RVM) to detect abnormalities in LUS images. Specifically, through pre-processing by an adaptive median filter and feature extraction, a complete local binary pattern mask was generated. Then, the binary RVM decides whether the given input LUS image corresponds to a healthy or unhealthy subject. If abnormal, the images will then be classified by the multi-level RVM into one of the lung diseases considered. This method finally leads to a classification accuracy of above 90\% and a 100\% specificity. 

\subsubsection{Transfer Learning}\label{subsec:4:2:3}
Transfer learning (TL) in medical tasks helps deep learning models achieve better performances when there are scarce medical images, and can also be defined as a solution to reducing training time. Horry et. al. \cite{2020COVID19} compared the performance of 8 widespread network structures, which are amenable to transfer learning for COVID-19 inspection with 3 modalities, including X-ray, CT and LUS. The results showed that their DL models are suitable for performing contrast enhancement on LUS images. The VGG19 outperforms all other models. The work in \cite{cheng2021transfer} used two pre-trained V-Unet and X-Unet models on 200 and 400 images respectively. In the former structure, the last three fully connected layers of a typical symmetrical U-Net structure were excluded and new layers for the expanding path were added, so that the VGG16 can be mimicked. The expanding path works reversely to the VGG16, with up-sampling through deconvolution by transposing convolutional layers. Then the V-Unet was pre-trained using ImageNet. The X-Unet is constructed by a U-Net pre-trained with grayscale salient object images. Almeida et. al. \cite{almeida2020lung} proposed the use of MobileNets (light-weight neural networks) to support the diagnosis of COVID-19. MobileNet was also pretrained on ImageNet, substituting the last layer for a fully connected layer with a ReLu activation and the output layer with a softmax activation. Multitask learning was achieved by training different MobileNet models for each pathological indication. Even with limited training data, the MobileNet achieved accuracy values above 95\% for all pneumonia indications. 

Compared with pre-training on real-life objects, for example ImageNet, the pre-training on LUS images may help detecting specific lung patterns such as B-Lines. Born et. al. \cite{born2020pocovidnet} published the first dataset of lung POCUS recordings of COVID-19, pneumonia, and healthy patients. The collected data is heterogeneous but was pre-processed manually to remove artifacts and checked by a medical doctor for quality. They also proposed a model based on a pre-trained CNN (POCOVID-Net) on the available data and evaluated using 5-fold cross validation. The POCOVID-NET contains a convolutional part in the form of VGG16 followed by one hidden layer with ReLu activation, and batch normalization followed by the output layer with SoftMax activation. The results reveal the efficiency in detecting COVID-19 by reaching a sensitivity of 0.96, a specificity of 0.79, and F1-score of 0.92.

Xue et. al. \cite{xue2021modality} introduced the idea of contrastive learning into severity assessment through LUS images, which falls under the transfer learning umbrella. The framework includes two modules: (i) a dual-level supervised multiple instance learning module (DSA-MIL) that leverages supervision from both the LUS zone scores and the patient severity of pneumonia; and (ii) a modality alignment contrastive learning module (MA-CLR) that combines representations of LUS and clinical information without dropping discriminative features. A staged representation transfer (SRT) strategy was introduced to train the nonlinear mapping, leveraging the semantic and discriminative information from the training data, providing an accuracy of 75\% for 4-way severity assessment and 87.5\% for two-way classification.

Panicker et. al. \mbox{\cite{panicker2021approach}} proposed an U-net based network structure called LUSNet and trained it in a similar way to contrastive learning. Fused images were used to train the LUSNet to learn the pleural region of interest in an unsupervised manner, and then the LUSNet was trained in a supervised manner to classify the images into five classes of increasing severity of infection. Images used for unsupervised training were fused with gray scale images by exploiting acoustic propagation features.This technique has the potential to improve the performance of the neural network when the input dataset is limited and diverse. Experiments on over 5000 frames of COVID-19 videos showed an average five-fold cross-validation accuracy, sensitivity, and specificity of 97\%, 93\%, and 98\% respectively.

\begin{figure}[htbp]
\centering
\includegraphics[width=.9\linewidth]{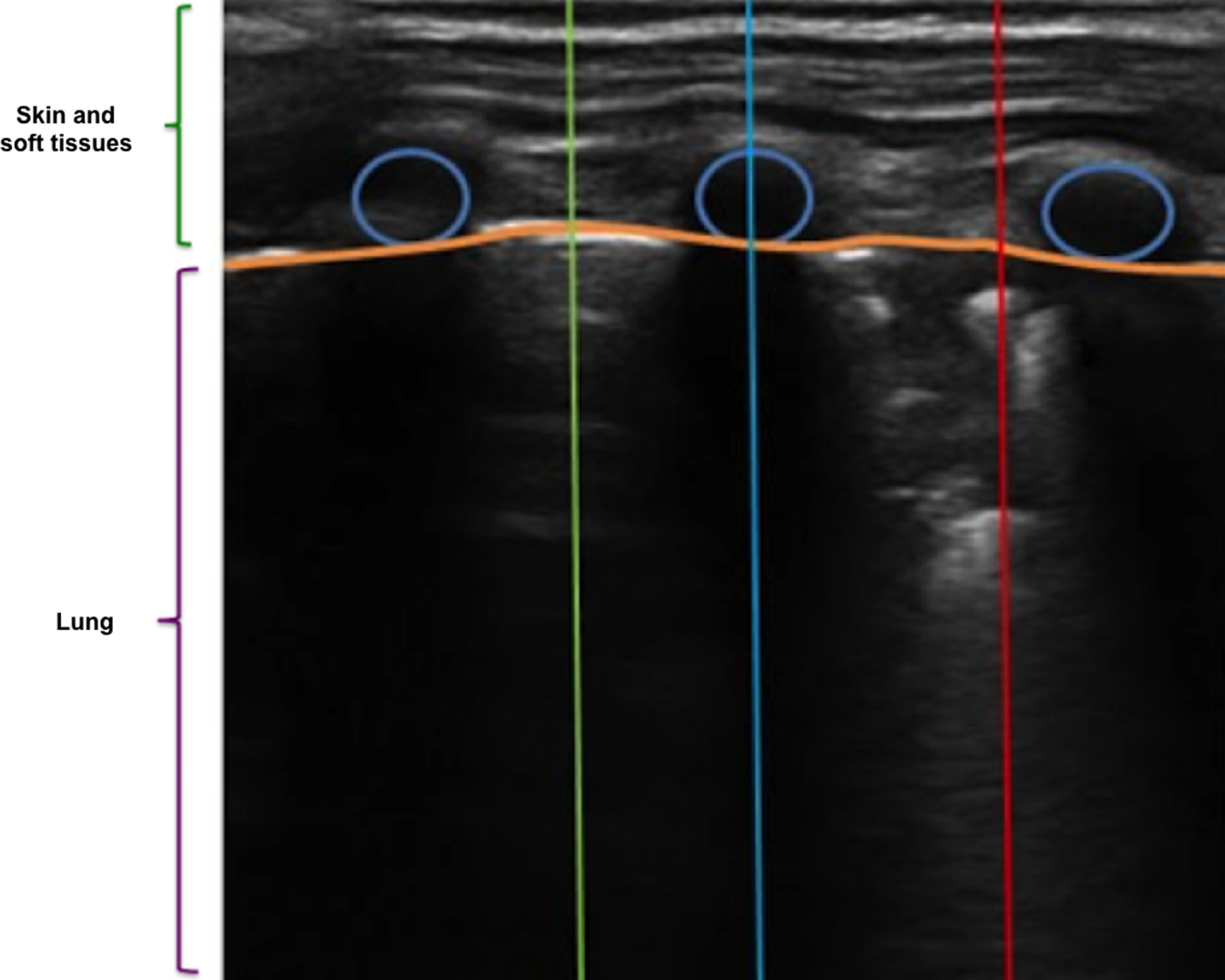}
\caption{Examples of vectors in specific regions used to compute brightness profiles: healthy (green), rib-bone (blue), and pneumonia
(red). \cite{2018Automatic}.}
\label{Malena_Automatic}
\end{figure}

\begin{figure}[htbp]
\centering
\includegraphics[width=.9\linewidth]{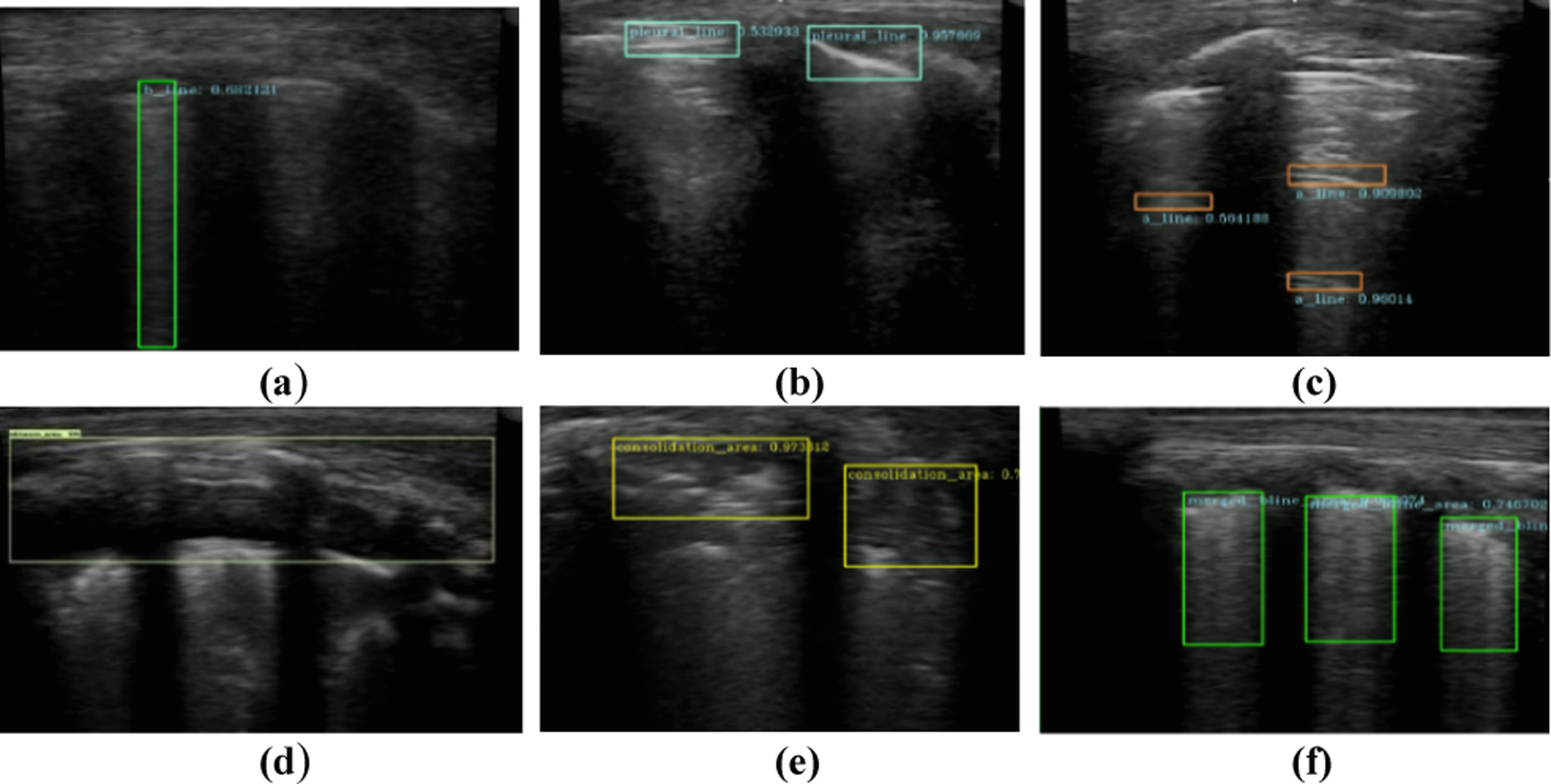}
\caption{Sample results for SSD detection models results of  Kulhare et. al. \cite{kulhare2018ultrasound}. (a) B-line, (b) pleural line, (c) A-line, (d) pleural effusion, (e) consolidation, (f) merged B-line.}
\label{Sourabh_USbased}
\end{figure}

\begin{figure}[htbp]
\centering
\includegraphics[width=.9\linewidth]{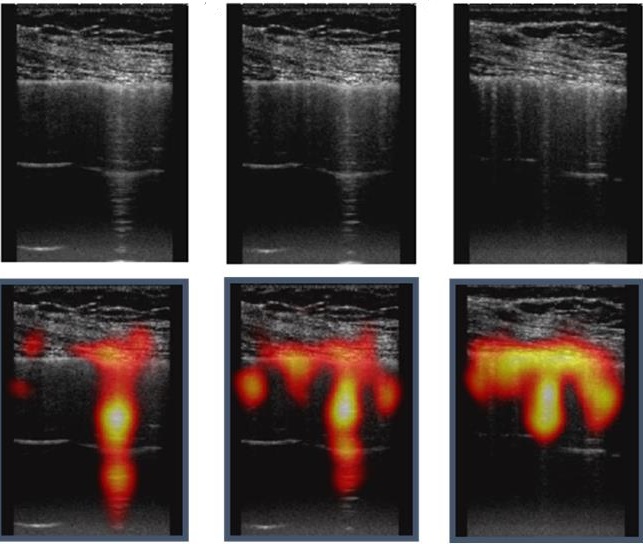}
\caption{B-line localization results in the work of \cite{2019Localizing}. Top: clinical B-mode input data. Bottom: corresponding class activation maps (CAM) for frames containing multiple B-lines.}
\label{ruud_localizing}
\end{figure}

\begin{figure}[htbp]
\centering
\includegraphics[width=.9\linewidth]{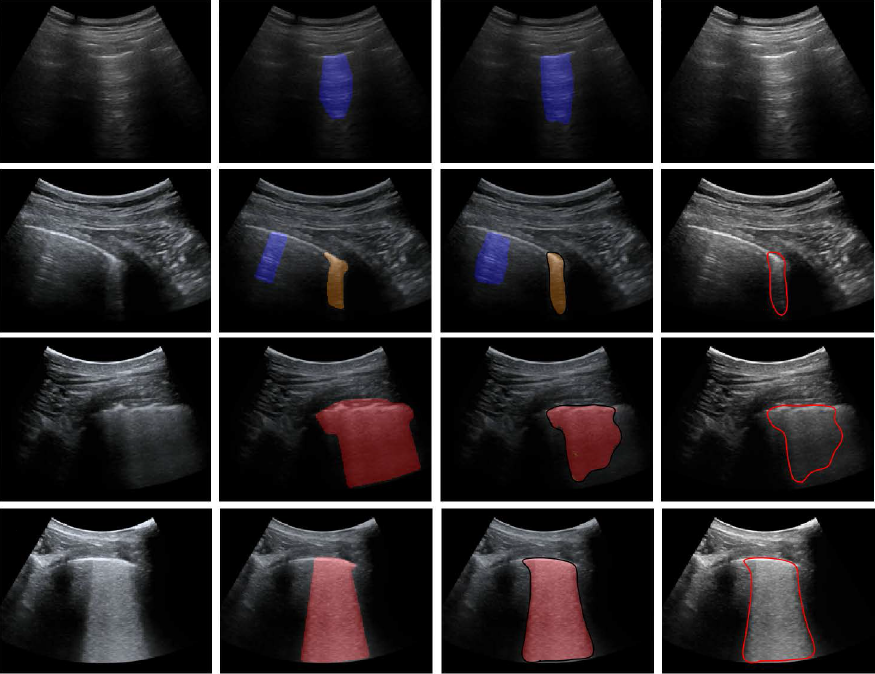}
\caption{Four examples of B-mode input image frames (first column), their annotations (second column), semantic segmentations (third column) and contours of COVID-19 markers by deep learning (forth column) \cite{2020DL4COVIDinPOC}.}
\label{Dorothy_TL}
\end{figure}

\begin{figure}[htbp]
\centering
\includegraphics[width=.92\linewidth]{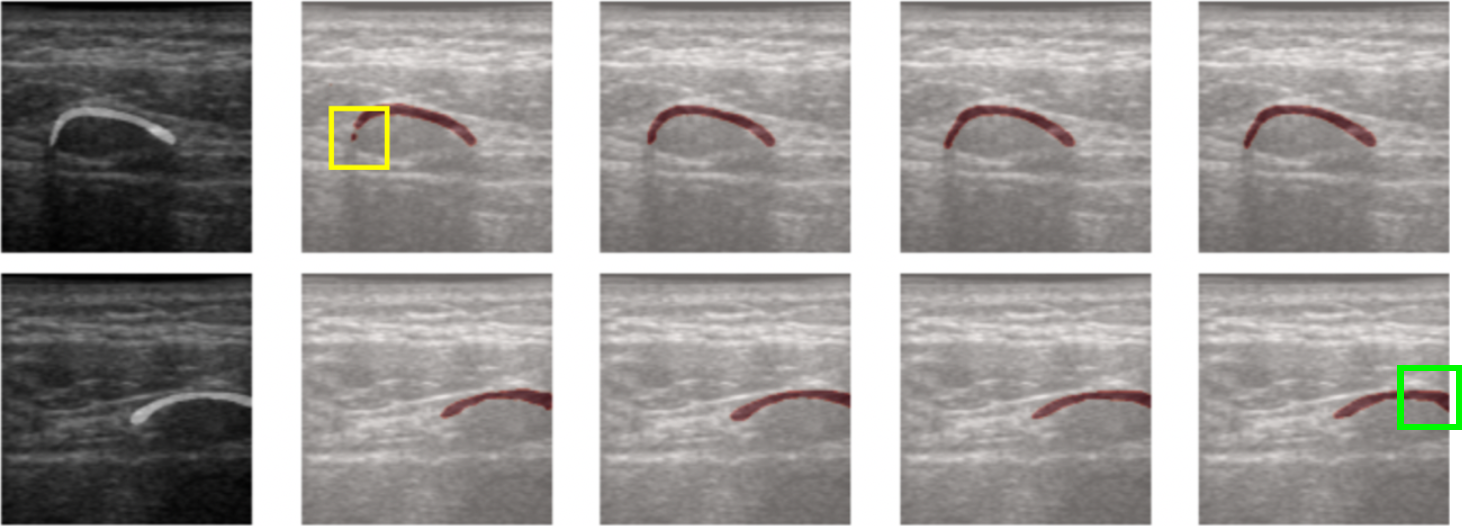}
\caption{Comparison of two examples from the LUS data between the original mask, predicted masks from V-Unet TL, V-Unet FT, X-Unet FT (All layers), and X-Unet FT (BBFrozen). Yellow and green boxes indicate the defects from V-Unet and XUnet respectively  \cite{cheng2021transfer}.}
\label{Subhankar_DL}
\end{figure}

\begin{figure}[htbp]
\centering
\includegraphics[width=.92\linewidth]{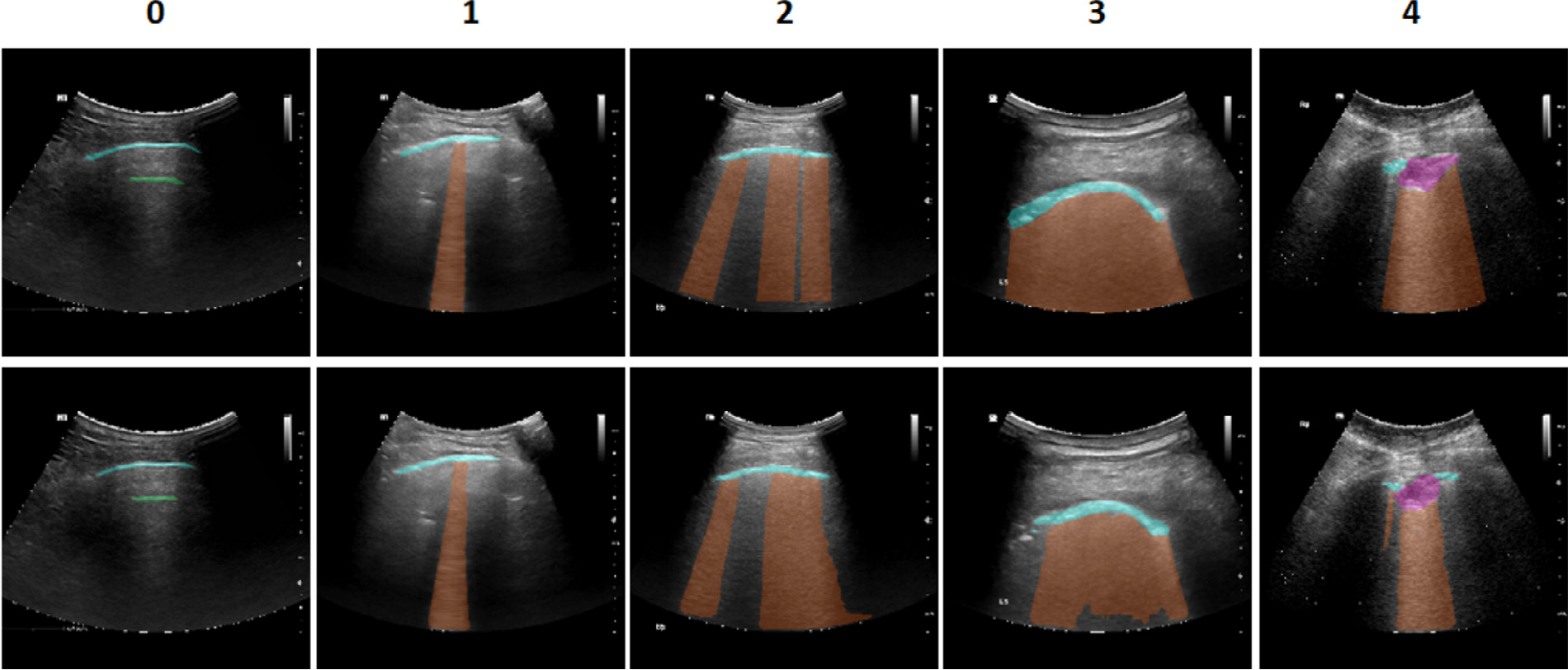}
\caption{Segmentation examples for a representative frame from five lung zones with different severity. Upper: ground truth, Lower: prediction, Cyan: pleural line, Green: A-line, Brown: B-line, Purple: consolidation \cite{xue2021modality}.}
\label{Wufeng_Modality}
\end{figure}

\begin{table*}[ht]
\scriptsize
  \centering
  \caption{Computational Techniques for LUS Image Analysis}\setstretch{0.9}\scriptsize
    \begin{tabular}{@{}|p{1.5cm}|p{2.5cm}|p{5.5cm}|p{5.8cm}|@{}} 
    \toprule
    Category & Approaches & Contributions & Limitations \\
    \toprule
    \multicolumn{1}{|c|}{\multirow{5}[4]{*}{Model-based}} & Alternate sequential filtering and top-hat filtering \cite{Ramin2018Automatic} & Successfully differentiate between the healthy and unhealthy class with a p-value of 0.015. Average number of B-lines of patients and healthy group are 0.28$\pm$0.06 and 0.03$\pm$0.06 respectively. & Lack the evaluation on larger dataset with different degrees of edema; Non real-time; No quantification considered. \\
\cline{2-4}          & Simple linear regression (SLR) and robust linear regression (RLR) \cite{brusasco2019quantitative} & Provides a reliable operator-independent assessment of EVLW in ARDS, which is comparable or superior to previous scores. &  Not suitable for the case when multiple  B-lines  coalesce  into  a single white line.  \\

\cline{2-4}          & Hidden Markov model, Viterbi algorithm  \cite{2020Automatic} & Average overall accuracy in detecting the pleura is 84\% and 94\% for convex and linear probes, respectively. & -  \\
\cline{2-4}          & Radon transform based inverse problem solved by ADMM \cite{2016Line,2017Line} & Improves the performance of B-line detection approaches by up to 50\%. & Non real-time. \\
\cline{2-4}          & Radon transform based non-convex optimization with Cauchy-based penalty function \cite{2020Detection} & Up to 88\% detection accuracy on 12-segment LUS for 9 COVID-19 patients.  & Non real-time. \\
    \hline
    \multicolumn{1}{|c|}{\multirow{14}[4]{*}{Data-driven}} & Characteristic vector classifiers \cite{2018Automatic} & 90.9\% Sensitivity and 100\% specificity for pneumonia infiltrates identification. & Only severe and radiographically evident cases are analyzed; The data only come from a single patient; The algorithm requires validation for individual ultra-sonographers.   \\
\cline{2-4}          & A  basis  CNN  with  data  augmentation  and dropout to avoid over-fitting \cite{2019Quantifying} & Requires littele computing power; Low number of parameters; Nearly real-time; Adaptive to other clinical tasks. & Lack comet-rich samples; Labeled by a single observer; Manually count the number of B-lines. \\
\cline{2-4}          & 3D-CsNet \cite{Cristiana2020Automated} & A small number of trainable parameters; Effective in poorer quality datasets. & Deficient categories of the training data leads to a weaker performance on multiclass severity rating compared to binary rating. \\
\cline{2-4}          & Single shot neural net \cite{kulhare2018ultrasound} & Light-weight network structure makes it suitable for on-device inference tasks.   & The application to human LUS images has not been investigated. \\
\cline{2-4}          & multilayer FNNs \cite{chen2021quantitative} &  The work is not dependent on arbitrary or human-decided thresholds and is capable of getting robust classification result based on limited training data. & The results are evaluated on visual determination of the B-score and no artefacts are considered other than B-lines. \\
\cline{2-4}          & CNN   for   B-line   detection   and   weakly-supervised  localization  through  CAM \cite{2019Localizing} & Nearly real-time; the algorithm can also serve as a region-of-interest selector for further quantitative analysis of segments. & The characterization of adequate phenotyping of various lung pathologies is needed.\\
\cline{2-4}          & RNN \cite{kerdegari2021b} & Detect the B-line artefacts and localize them both spatially and temporally within LUS videos. & Focused on B-line artefacts only. \\
\cline{2-4}          & STN  based  deep  network \cite{2020DL4COVIDinPOC} & Extended and fully-annotated version of the ICLUS-DB database. & The precise demographics of the patient group in the database remain unknown; The possible inclusion of false positive cases; Noisy labels in the database.\\
\cline{2-4}          & multi-level RVM \cite{2015Detection} & Classification accuracy of above 90\% and 100\% specificity when tested on the dataset. & The size of the dataset is not known.  \\
\cline{2-4}          & SVM \cite{2020Automatic} & The need of small dataset by using SVM.  & The learning stage is time-consuming. \\
\cline{2-4}          & V-Unet and X-Unet \cite{cheng2021transfer} & Short training time; Accurate when scare medical images.  & No regularization may lead to over-fitting to US images only; Performance is dataset-dependent.\\
\cline{2-4}          & POCOVID-Net \cite{born2020pocovidnet} & The first dataset of lung POCUS recordings of COVID-19, pneumonia, and healthy patients.  & The dataset is restricted to convex probe. \\
\cline{2-4}          & A  DSA-MIL  module  to  predict  the  patentseverity  and  a  MA-CLR  for  combination  ofthe LUS data and the clinical information \cite{xue2021modality} & Issues related to data heterogeneity, multi-modality and non-linearity are alleviated. & The distribution of lung zone information is ignored. \\
\cline{2-4}          & LUSNet \cite{panicker2021approach} &  Improves the performance of the neural network significantly and also aids the labelling by combining acoustic propagation features with gray scale images. & Restricted only to COVID-19. \\
    \hline
    \end{tabular}%
    \label{tab:Methodology}%
    \end{table*}%

\subsection{Model-based vs Data-driven LUS Image Analysis}\label{subsec:3:4}

Table \ref{tab:Methodology} provides a summary of all the aforementioned methods. Fig \ref{Ramin_automatic} to Fig \ref{Wufeng_Modality} show some example results from the reviewed literature.

For the model-based approaches, the most obvious advantage is the lack of need for training, and that a large dataset is not required. Given a specific task, the model-based methods are able to show good performances in terms of accuracy as the feature extraction algorithms are purposely designed by experts. Some rapid-converging algorithms enable the analysis of LUS image sequences in a relatively short amount of time, but it is still hard to achieve real-time processing. Meanwhile, being inherently explainable, the effects of the algorithms can be intuitively understood in each processing step, so they are straightforward to modify. 

However, considering the task-oriented nature of the model-based methods, the ability of generalization can be limited in complex clinical cases. Therefore, data-driven approaches are investigated. In DL methods, network parameters need training through a large amount of data, so as to achieve the purposes of object detection, recognition, segmentation and so on. When enough labeled data are available, DL can achieve or even exceed the performance of humans \cite{HEKLER201991}. The supervised learning approaches have reached the accuracy required for clinical diagnosis \cite{supriya2020machine}. However, for LUS, it is usually difficult to acquire a large number of multi-category data, so attempts to design weakly-supervised and transfer learning methods have been made. For transfer learning, the network trained with LUS images will have a better performance than that trained with  natural images \cite{born2020pocovidnet}. In clinical scenarios, the high detection efficiency is a must, and the data processing speed should be fast. Therefore, real-time tasks and computationally efficient ML approaches have also been studied.

\section{\label{sec:5} Challenges and Future Research Directions}

From the literature reviewed above, the importance of LUS in assisting clinical diagnosis has been demonstrated thoroughly. However, although LUS has a proven capability to intercept alterations along the lung surface and value in treatment monitoring, it is still facing challenges in real application.

Despite the fact that model-based methods can fulfil the aim of line identification, their generalization ability is not as good as data-driven approaches such as ML techniques. But in future researches, one should not neglect the inherent interpretability of the classical algorithms, which may help alleviate the black box problem of ML/DL techniques. 

As the application of data-driven approaches is predominant in the medical field, data standardization becomes an essential problem. This concept concerns various aspects of LUS image processing, such as data collection, data labelling, and algorithm validation etc. So far, there is no regularization in LUS image collection. Therefore the quality and the form of the images are dependent on the equipment and the operator. This inconsistency in data collection along with the lack of labeling standards brings difficulties in providing an accurate ground truth of the lung condition, because the diagnosis results varies among clinicians. Without a well-labelled dataset to rely on, the output of the trained network would not be trusty enough in clinical cases. Accordingly, gold rules for setting up standardized datasets are urgently needed. What's more, to maximize working efficiency, the requirement on real-time artefact detection is also highly expected by clinicians.

Even though the challenges mentioned above have not been adequately tackled yet, several prospective data-driven techniques in medical image processing are worth studying. For example, transfer learning is a feasible solution to the lack of suitable datasets. A pretrained network can be first obtained on ImageNet, and then be used for LUS image training. However the performance has not been demonstrated to be as good as targeted training on medical image datasets \cite{DLinMUS}.  More recently, looking at current trends in the ML community with respect to DL, we identify as key area which can be relevant for LUS image processing that of self-supervised learning. Different from fully supervised learning, self-supervised learning does not require manual labeling since it generates them by itself \cite{lecun2021self}. In this sense, self-supervised learning can be considered as a subset of unsupervised learning methods. Self-supervised learning methods are attractive as they have been able to utilize unlabelled data to learn the underlying representations. This is where self-supervised methods play a vital tole in fuelling the progress of DL without the need for expensive annotations and learn feature representations where data provide supervision \cite{jing2020self}. The successful application of self-supervised learning in MRIs \cite{jamaludin2017self} sets a hope in its usage in other medical modalities including LUS. Moreover, by compacting light-weigh network structures, the computational efficiency will be possibly further improved.

Computational approaches have the potential to make quantitative LUS assessment faster and more reliable, particularly in nonexpert hands \mbox{\cite{mongodi2021quantitative}}. Automatic detection, localization and quantification of B-lines are tasks of great clinical interest and huge potential for clinical applications. Building upon the idea of quantifying B-lines with neural networks \mbox{\cite{chen2021quantitative}}, hybrid (i.e. combined model-based and data-driven) quantitative LUS approaches will open up new prospects for use not only as a diagnostic but also as a monitoring tool.

\section{\label{sec:6} Conclusions}

Clinical applications of LUS have become more and more numerous. In this article, we reviewed recent research progress in computational LUS image reconstruction and analysis, including model-based methods as well as data-driven methods. The approaches presented in this article all show the possibility of assisting disease diagnosis by detecting and quantifying line structures, especially through B-lines. The model-based methods are unsupervised and explainable, whilst the data-driven methods provide better generalization and more rapid development. In future studies, the combination of the two methodologies can be a trend, as a result of which the requirement of labelled dataset can be relaxed while keeping the performances of the algorithms. Furthermore, the availability of the LUS dataset is improving, which will likely nurture to algorithmic development. 

\bibliographystyle{IEEEtran}
\bibliography{ReviewRef.bib}

\begin{thebibliography}{10}
\providecommand{\url}[1]{#1}
\csname url@samestyle\endcsname
\providecommand{\newblock}{\relax}
\providecommand{\bibinfo}[2]{#2}
\providecommand{\BIBentrySTDinterwordspacing}{\spaceskip=0pt\relax}
\providecommand{\BIBentryALTinterwordstretchfactor}{4}
\providecommand{\BIBentryALTinterwordspacing}{\spaceskip=\fontdimen2\font plus
\BIBentryALTinterwordstretchfactor\fontdimen3\font minus
  \fontdimen4\font\relax}
\providecommand{\BIBforeignlanguage}[2]{{%
\expandafter\ifx\csname l@#1\endcsname\relax
\typeout{** WARNING: IEEEtran.bst: No hyphenation pattern has been}%
\typeout{** loaded for the language `#1'. Using the pattern for}%
\typeout{** the default language instead.}%
\else
\language=\csname l@#1\endcsname
\fi
#2}}
\providecommand{\BIBdecl}{\relax}
\BIBdecl

\bibitem{allinovi2017lung}
M.~Allinovi, M.~Saleem, P.~Romagnani, P.~Nazerian, and W.~Hayes, ``Lung
  ultrasound: a novel technique for detecting fluid overload in children on
  dialysis,'' \emph{Nephrology Dialysis Transplantation}, vol.~32, no.~3, pp.
  541--547, 2017.

\bibitem{touw2015lung}
H.~Touw, P.~Tuinman, H.~Gelissen, E.~Lust, and P.~Elbers, ``Lung ultrasound:
  routine practice for the next generation of internists,'' \emph{Neth J Med},
  vol.~73, no.~3, pp. 100--107, 2015.

\bibitem{touw2019routine}
H.~Touw, A.~E. Schuitemaker, F.~Daams, D.~L. van~der Peet, E.~Bronkhorst,
  P.~Schober, C.~Boer, and P.~R. Tuinman, ``Routine lung ultrasound to detect
  postoperative pulmonary complications following major abdominal surgery: a
  prospective observational feasibility study,'' \emph{The Ultrasound Journal},
  vol.~11, no.~1, pp. 1--8, 2019.

\bibitem{FIRS}
E.~R. Society, ``Forum of international respiratory societies. the global
  impact of respiratory disease – second edition.''
  \url{https://www.who.int/gard/publications/The_Global_Impact_of_Respiratory_Disease.pdf},
  2017.

\bibitem{BLungFoundation}
\BIBentryALTinterwordspacing
BLF.org, ``Lung disease in the {UK},'' 2021. [Online]. Available:
  \url{https://statistics.blf.org.uk/?_ga=2.18220605.520238884.1611621128-714148984.1611621128}
\BIBentrySTDinterwordspacing

\bibitem{LUfuture}
L.~Demi, ``Lung ultrasound: The future ahead and the lessons learned from
  covid-19,'' \emph{The Journal of the Acoustical Society of America}, vol.
  148, no.~4, pp. 2146--2150, 2020.

\bibitem{Clinical}
Z.~Hua, \emph{Clinical Analysis of Ultrasound Diagnosis of PULMONARY
  DISEASES}.\hskip 1em plus 0.5em minus 0.4em\relax Beijing: Peking University
  Medical Press, 2019.

\bibitem{2016Inverse}
N.~Zhao, ``Inverse problems in medical ultrasound images - applications to
  image deconvolution, segmentation and super-resolution,'' Ph.D. dissertation,
  Institut National Polytechnique de Toulouse, 2016.

\bibitem{USimagingmodel}
J.~{Ng}, R.~{Prager}, N.~{Kingsbury}, G.~{Treece}, and A.~{Gee}, ``Modeling
  ultrasound imaging as a linear, shift-variant system,'' \emph{IEEE
  Transactions on Ultrasonics, Ferroelectrics, and Frequency Control}, vol.~53,
  no.~3, pp. 549--563, 2006.

\bibitem{2019Application2}
R.~Raheja, M.~Brahmavar, D.~Joshi, and D.~Raman, ``Application of lung
  ultrasound in critical care setting: A review,'' \emph{Cureus}, vol.~11,
  no.~7, 2019.

\bibitem{volpicelli2013lung}
G.~Volpicelli, ``Lung sonography,'' \emph{Journal of Ultrasound in Medicine},
  vol.~32, no.~1, pp. 165--171, 2013.

\bibitem{gargani2014lung}
L.~Gargani and G.~Volpicelli, ``How i do it: lung ultrasound,''
  \emph{Cardiovascular ultrasound}, vol.~12, no.~1, pp. 1--10, 2014.

\bibitem{MedicalEquip}
Z.~Qiang, \emph{Medical Ultrasonic Imaging Equipment}.\hskip 1em plus 0.5em
  minus 0.4em\relax Shanghai: The Second Military Medical University Press,
  2000.

\bibitem{carroll2021mmode}
\BIBentryALTinterwordspacing
D.~Carroll, ``M-mode (ultrasound) | radiology reference article |
  radiopaedia.org,'' Radiopaedia, 11 2021. [Online]. Available:
  \url{https://radiopaedia.org/articles/m-mode-ultrasound}
\BIBentrySTDinterwordspacing

\bibitem{USbasis}
B.~Slak, ``Development, optimization and clinical evaluation of algorithms for
  ultrasound data analysis used in selected medical applications,'' \url{
  https://scholar.uwindsor.ca/etd/8339}, 2020.

\bibitem{soldati2020proposal}
G.~Soldati, A.~Smargiassi, R.~Inchingolo, D.~Buonsenso, T.~Perrone, D.~F.
  Briganti, S.~Perlini, E.~Torri, A.~Mariani, E.~E. Mossolani \emph{et~al.},
  ``Proposal for international standardization of the use of lung ultrasound
  for patients with covid-19: a simple, quantitative, reproducible method,''
  \emph{Journal of Ultrasound in Medicine}, vol.~39, no.~7, pp. 1413--1419,
  2020.

\bibitem{allinovi2021simplified}
\BIBentryALTinterwordspacing
M.~Allinovi and W.~Hayes, ``{Simplified 8-site lung ultrasound examination to
  assess fluid overload in children on haemodialysis},'' \emph{Clinical Kidney
  Journal}, 02 2021, sfab041. [Online]. Available:
  \url{https://doi.org/10.1093/ckj/sfab041}
\BIBentrySTDinterwordspacing

\bibitem{soldati2015lung}
G.~Soldati, A.~Smargiassi, R.~Inchingolo, S.~Sher, R.~Nenna, S.~Valente, and
  C.~D. Inchingolo, ``Lung ultrasonography and vertical artifacts: The shape of
  air.'' \emph{Respiration}, vol.~90, no.~1, pp. 86--87, 2015.

\bibitem{van2019b}
R.~J. van Sloun and L.~Demi, ``B-line detection and localization by means of
  deep learning: preliminary in-vitro results,'' in \emph{International
  Conference on Image Analysis and Recognition}.\hskip 1em plus 0.5em minus
  0.4em\relax Springer, 2019, pp. 418--424.

\bibitem{Wang2021}
Y.~Wang, Y.~Zhang, Q.~He, H.~Liao, and J.~Luo, ``Pleural line and b-lines based
  image analysis for severity evaluation of covid-19 pneumonia,'' in \emph{2021
  IEEE International Ultrasonics Symposium (IUS)}, 2021, pp. 1--4.

\bibitem{Ramin2018Automatic}
Ramin, Moshavegh, K.~Lindskov, Hansen, Hasse, Moller-Sorensen, M.~Bachmann,
  Nielsen, J.~Arendt, and Jensen, ``Automatic detection of b-lines in in-vivo
  lung ultrasound.'' \emph{IEEE Transactions on Ultrasonics, Ferroelectrics,
  and Frequency Control}, 2018.

\bibitem{2019DEVICES}
L.~Hu, C.~Mehanian, B.~K. Wilson, and X.~Zheng, ``Devices, systems, and methods
  for diagnosis of pulmonary conditions through detection of b-lines in lung
  sonography,'' 2019.

\bibitem{2020Automatic}
C.~Leonardo, D.~Elena, M.~Daniele, Z.~Massimo, M.~Federico, T.~Elena,
  S.~Andrea, I.~Riccardo, S.~Gino, D.~Libertario, B.~Francesca, and B.~Lorenzo,
  ``Automatic pleural line extraction and covid-19 scoring from lung ultrasound
  data,'' \emph{IEEE Transactions on Ultrasonics, Ferroelectrics, and Frequency
  Control}, vol.~67, no.~11, pp. 2207--2217, 2020.

\bibitem{2018Automatic}
M.~Correa, M.~Zimic, F.~Barrientos, R.~Barrientos, A.~Román-Gonzalez,
  M.~Pajuelo, C.~Anticona, H.~Mayta, A.~Alva, L.~Solis-Vasquez, D.~Figueroa,
  M.~Chavez, R.~Lavarello, B.~Castañeda, V.~Paz-Soldán, W.~Checkley,
  R.~Gilman, and R.~Oberhelman, ``Automatic classification of pediatric
  pneumonia based on lung ultrasound pattern recognition.'' \emph{PLoS ONE},
  2018.

\bibitem{Lichtenstein2016Lung}
D.~A. Lichtenstein, ``Lung ultrasound in the critically ill,'' \emph{Annals of
  intensive care}, vol.~4, no.~1, pp. 1--12, 2014.

\bibitem{2019Quantifying}
X.~Wang, J.~S. Burzynski, J.~Hamilton, P.~S. Rao, W.~F. Weitzel, and J.~L.
  Bull, ``Quantifying lung ultrasound comets with a convolutional neural
  network: Initial clinical results,'' \emph{Computers in Biology and
  Medicine}, 2019.

\bibitem{allinovi2016finding}
M.~Allinovi, M.~A. Saleem, O.~Burgess, C.~Armstrong, and W.~Hayes, ``Finding
  covert fluid: methods for detecting volume overload in children on
  dialysis,'' \emph{Pediatric Nephrology}, vol.~31, no.~12, pp. 2327--2335,
  2016.

\bibitem{rippey2020lung}
\BIBentryALTinterwordspacing
J.~Rippey, ``Lung ultrasound technique - overview,'' Life in the Fast Lane, 11
  2020. [Online]. Available:
  \url{https://litfl.com/lung-ultrasound-technique-overview/}
\BIBentrySTDinterwordspacing

\bibitem{2020What}
G.~Volpicelli, A.~Lamorte, and T.~Villén, ``What's new in lung ultrasound
  during the covid-19 pandemic,'' \emph{Intensive Care Medicine}, vol.~46,
  no.~4, 2020.

\bibitem{2020Point}
M.~J. Smith, S.~A. Hayward, S.~M. Innes, and A.~S.~C. Miller,
  ``Point‐of‐care lung ultrasound in patients with covid‐19 – a
  narrative review,'' \emph{Anaesthesia}, 2020.

\bibitem{soldati2020there}
G.~Soldati, A.~Smargiassi, R.~Inchingolo, D.~Buonsenso, T.~Perrone, D.~F.
  Briganti, S.~Perlini, E.~Torri, A.~Mariani, E.~E. Mossolani \emph{et~al.},
  ``Is there a role for lung ultrasound during the covid-19 pandemic?''
  \emph{Journal of Ultrasound in Medicine}, 2020.

\bibitem{zhou2020ultrasound}
B.~Zhou, X.~Yang, X.~Zhang, W.~J. Curran, and T.~Liu, ``Ultrasound elastography
  for lung disease assessment,'' \emph{IEEE Transactions on Ultrasonics,
  Ferroelectrics, and Frequency Control}, vol.~67, no.~11, pp. 2249--2257,
  2020.

\bibitem{2018Wu}
S.~Wu, ``Advances in the clinical application of lung ultrasonography,''
  \emph{Advances in Clinical Medicine}, vol.~8, no.~7, 2018.

\bibitem{Eldar2005RMSE}
Y.~C. {Eldar}, A.~{Ben-Tal}, and A.~{Nemirovski}, ``Robust mean-squared error
  estimation in the presence of model uncertainties,'' \emph{IEEE Transactions
  on Signal Processing}, vol.~53, no.~1, pp. 168--181, 2005.

\bibitem{Achim2001Novel}
A.~{Achim}, A.~{Bezerianos}, and P.~{Tsakalides}, ``Novel bayesian multiscale
  method for speckle removal in medical ultrasound images,'' \emph{IEEE
  Transactions on Medical Imaging}, vol.~20, no.~8, pp. 772--783, 2001.

\bibitem{mohanty2020vivo}
K.~Mohanty, Y.~Karbalaeisadegh, J.~Blackwell, M.~Ali, B.~Masuodi, T.~Egan, and
  M.~Muller, ``In vivo assessment of pulmonary fibrosis and pulmonary edema in
  rodents using ultrasound multiple scattering,'' \emph{IEEE Transactions on
  Ultrasonics, Ferroelectrics, and Frequency Control}, vol.~67, no.~11, pp.
  2274--2280, 2020.

\bibitem{lye2021vivo}
T.~H. Lye, R.~Roshankhah, Y.~Karbalaeisadegh, S.~A. Montgomery, T.~M. Egan,
  M.~Muller, and J.~Mamou, ``In vivo assessment of pulmonary fibrosis and edema
  in rodents using the backscatter coefficient and envelope statistics,''
  \emph{The Journal of the Acoustical Society of America}, vol. 150, no.~1, pp.
  183--192, 2021.

\bibitem{2020Despeckling}
H.~Choi and J.~Jeong, ``Despeckling algorithm for removing speckle noise from
  ultrasound images,'' \emph{Symmetry}, vol.~12, no.~6, pp. 1--26, 2020.

\bibitem{Yong2002SRAD}
{Yongjian Yu} and S.~T. {Acton}, ``Speckle reducing anisotropic diffusion,''
  \emph{IEEE Transactions on Image Processing}, vol.~11, no.~11, pp.
  1260--1270, 2002.

\bibitem{2020chen}
B.~Chen, Y.~Lv, J.~Zou, W.~Chen, and B.~Pan, ``A novel speckle noise removal
  algorithm based on admm and energy minimization method,'' \emph{Journal of
  Function Spaces}, vol. 2020, no.~7, pp. 1--17, 2020.

\bibitem{Boyd2011Alternating}
S.~Boyd, N.~Parikh, and E.~Chu, \emph{Distributed optimization and statistical
  learning via the alternating direction method of multipliers}.\hskip 1em plus
  0.5em minus 0.4em\relax Now Publishers Inc, 2011.

\bibitem{jensen1993deconvolution}
J.~A. Jensen, J.~Mathorne, T.~Gravesen, and B.~Stage, ``Deconvolution of
  in-vivo ultrasound b-mode images,'' \emph{Ultrasonic Imaging}, vol.~15,
  no.~2, pp. 122--133, 1993.

\bibitem{Hourani2020}
M.~Hourani, A.~Basarab, F.~Varray, D.~Kouamé, and J.-Y. Tourneret,
  ``Block-wise ultrasound image deconvolution from fundamental and harmonic
  images,'' in \emph{2020 IEEE International Ultrasonics Symposium (IUS)},
  2020, pp. 1--4.

\bibitem{Hourani2021}
M.~Hourani, A.~Basarab, D.~Kouamé, and J.-Y. Tourneret, ``Ultrasound image
  deconvolution using fundamental and harmonic images,'' \emph{IEEE
  Transactions on Ultrasonics, Ferroelectrics, and Frequency Control}, vol.~68,
  no.~4, pp. 993--1006, 2021.

\bibitem{taxt1995restoration}
T.~Taxt, ``Restoration of medical ultrasound images using two-dimensional
  homomorphic deconvolution,'' \emph{IEEE transactions on ultrasonics,
  ferroelectrics, and frequency control}, vol.~42, no.~4, pp. 543--554, 1995.

\bibitem{michailovich2005novel}
O.~V. Michailovich and D.~Adam, ``A novel approach to the 2-d blind
  deconvolution problem in medical ultrasound,'' \emph{IEEE transactions on
  medical imaging}, vol.~24, no.~1, pp. 86--104, 2005.

\bibitem{2011Statistical}
M.~Alessandrini, ``Statistical methods for analysis and processing of medical
  ultrasound: applications to segmentation and restoration,'' \emph{These de
  doctorat}, 2011.

\bibitem{besson2019physical}
A.~Besson, L.~Roquette, D.~Perdios, M.~Simeoni, M.~Arditi, P.~Hurley, Y.~Wiaux,
  and J.-P. Thiran, ``A physical model of nonstationary blur in ultrasound
  imaging,'' \emph{IEEE Transactions on Computational Imaging}, vol.~5, no.~3,
  pp. 381--394, 2019.

\bibitem{michailovich2019iterative}
O.~Michailovich, A.~Basarab, and D.~Kouame, ``Iterative reconstruction of
  medical ultrasound images using spectrally constrained phase updates,'' in
  \emph{2019 IEEE 16th International Symposium on Biomedical Imaging (ISBI
  2019)}.\hskip 1em plus 0.5em minus 0.4em\relax IEEE, 2019, pp. 1765--1768.

\bibitem{2020Joint}
D.~H. Pham, A.~Basarab, I.~Zemmoura, J.~P. Remenieras, and D.~Kouame, ``Joint
  blind deconvolution and robust principal component analysis for blood flow
  estimation in medical ultrasound imaging,'' \emph{IEEE transactions on
  ultrasonics, ferroelectrics, and frequency control}, 2020.

\bibitem{8808885}
R.~J.~G. {van Sloun}, R.~{Cohen}, and Y.~C. {Eldar}, ``Deep learning in
  ultrasound imaging,'' \emph{Proceedings of the IEEE}, vol. 108, no.~1, pp.
  11--29, 2020.

\bibitem{vedula2017ctquality}
S.~Vedula, O.~Senouf, A.~M. Bronstein, O.~V. Michailovich, and M.~Zibulevsky,
  ``Towards ct-quality ultrasound imaging using deep learning,'' 2017.

\bibitem{BookSpeckle}
D.~Feng, W.~Wu, H.~Li, and Q.~Li, \emph{Speckle Noise Removal in Ultrasound
  Images Using a Deep Convolutional Neural Network and a Specially Designed
  Loss Function}.\hskip 1em plus 0.5em minus 0.4em\relax Switzerland: Springer
  Nature Switzerland AG, 01 2020, pp. 85--92.

\bibitem{2017ADL2US}
D.~Perdios, A.~Besson, M.~Arditi, and J.-P. Thiran, ``A deep learning approach
  to ultrasound image recovery,'' in \emph{2017 IEEE International Ultrasonics
  Symposium (IUS)}.\hskip 1em plus 0.5em minus 0.4em\relax Ieee, 2017, pp.
  1--4.

\bibitem{2017DL4AccelUS}
Y.~H. Yoon and J.~C. Ye, ``Deep learning for accelerated ultrasound imaging,''
  in \emph{2018 IEEE International Conference on Acoustics, Speech and Signal
  Processing (ICASSP)}.\hskip 1em plus 0.5em minus 0.4em\relax IEEE, 2018, pp.
  6673--6676.

\bibitem{2016Deep}
K.~H. Jin, M.~T. Mccann, E.~Froustey, and M.~Unser, ``Deep convolutional neural
  network for inverse problems in imaging,'' \emph{IEEE Transactions on Image
  Processing}, vol.~PP, no.~99, pp. 4509--4522, 2016.

\bibitem{ronneberger2015u}
O.~Ronneberger, P.~Fischer, and T.~Brox, ``U-net: Convolutional networks for
  biomedical image segmentation,'' in \emph{International Conference on Medical
  image computing and computer-assisted intervention}.\hskip 1em plus 0.5em
  minus 0.4em\relax Springer, 2015, pp. 234--241.

\bibitem{2017One}
J.~H.~R. Chang, C.~L. Li, B.~Poczos, B.~V. K.~V. Kumar, and A.~C.
  Sankaranarayanan, ``One network to solve them all --- solving linear inverse
  problems using deep projection models,'' in \emph{Proceedings of the IEEE
  International Conference on Computer Vision}, 2017, pp. 5888--5897.

\bibitem{illpose}
Manifold, ``Ill-posed problem or inverse problem in image processing,'' \url{
  http://blog.sina.com.cn/s/blog_6833a4df0100nne9.html}, 2010.

\bibitem{2016Line}
N.~Anantrasirichai, M.~Allinovi, W.~Hayes, and A.~Achim, ``Automatic b-line
  detection in paediatric lung ultrasound,'' in \emph{2016 IEEE International
  Ultrasonics Symposium (IUS)}, 2016, pp. 1--4.

\bibitem{2017Line}
N.~Anantrasirichai, W.~Hayes, M.~Allinovi, D.~Bull, and A.~Achim, ``Line
  detection as an inverse problem: Application to lung ultrasound imaging,''
  \emph{IEEE Transactions on Medical Imaging}, vol.~PP, no.~99, pp. 1--1, 2017.

\bibitem{2020Detection}
O.~Karakus, N.~Anantrasirichai, A.~Aguersif, S.~Silva, A.~Basarab, and
  A.~Achim, ``Detection of line artefacts in lung ultrasound images of covid-19
  patients via non-convex regularization,'' \emph{IEEE Transactions on
  Ultrasonics Ferroelectrics and Frequency Control}, vol.~PP, no.~99, pp. 1--1,
  2020.

\bibitem{Karakus2020}
\BIBentryALTinterwordspacing
O.~Karakus, P.~Mayo, and A.~Achim, ``Convergence guarantees for non-convex
  optimisation with cauchy-based penalties,'' \emph{IEEE Transactions on Signal
  Processing}, vol.~68, p. 6159–6170, 2020. [Online]. Available:
  \url{http://dx.doi.org/10.1109/TSP.2020.3032231}
\BIBentrySTDinterwordspacing

\bibitem{brusasco2019quantitative}
C.~Brusasco, G.~Santori, E.~Bruzzo, R.~Tr{\`o}, C.~Robba, G.~Tavazzi,
  F.~Guarracino, F.~Forfori, P.~Boccacci, and F.~Corradi, ``Quantitative lung
  ultrasonography: a putative new algorithm for automatic detection and
  quantification of b-lines,'' \emph{Critical Care}, vol.~23, no.~1, pp. 1--7,
  2019.

\bibitem{roshankhah2021investigating}
R.~Roshankhah, Y.~Karbalaeisadegh, H.~Greer, F.~Mento, G.~Soldati,
  A.~Smargiassi, R.~Inchingolo, E.~Torri, T.~Perrone, S.~Aylward \emph{et~al.},
  ``Investigating training-test data splitting strategies for automated
  segmentation and scoring of covid-19 lung ultrasound images,'' \emph{The
  Journal of the Acoustical Society of America}, vol. 150, no.~6, pp.
  4118--4127, 2021.

\bibitem{Cristiana2020Automated}
C.~{Baloescu}, G.~{Toporek}, S.~{Kim}, K.~{McNamara}, R.~{Liu}, M.~M. {Shaw},
  R.~L. {McNamara}, B.~I. {Raju}, and C.~L. {Moore}, ``Automated lung
  ultrasound b-line assessment using a deep learning algorithm,'' \emph{IEEE
  Transactions on Ultrasonics, Ferroelectrics, and Frequency Control}, vol.~67,
  no.~11, pp. 2312--2320, 2020.

\bibitem{kulhare2018ultrasound}
S.~Kulhare, X.~Zheng, C.~Mehanian, C.~Gregory, M.~Zhu, K.~Gregory, H.~Xie,
  J.~M. Jones, and B.~Wilson, ``Ultrasound-based detection of lung
  abnormalities using single shot detection convolutional neural networks,'' in
  \emph{Simulation, Image Processing, and Ultrasound Systems for Assisted
  Diagnosis and Navigation}.\hskip 1em plus 0.5em minus 0.4em\relax Springer,
  2018, pp. 65--73.

\bibitem{McDermott2021Sonographic}
\BIBentryALTinterwordspacing
C.~McDermott, M.~Łącki, B.~Sainsbury, J.~Henry, M.~Filippov, and C.~Rossa,
  ``Sonographic diagnosis of covid-19: A review of image processing for lung
  ultrasound,'' \emph{Frontiers in Big Data}, vol.~4, p.~2, 2021. [Online].
  Available:
  \url{https://www.frontiersin.org/article/10.3389/fdata.2021.612561}
\BIBentrySTDinterwordspacing

\bibitem{chen2021quantitative}
J.~Chen, C.~Hef, J.~Yin, J.~Li, X.~Duan, Y.~Cao, L.~Sun, M.~Hu, W.~Lia, and
  Q.~Lib, ``Quantitative analysis and automated lung ultrasound scoring for
  evaluating covid-19 pneumonia with neural networks,'' \emph{IEEE Transactions
  on Ultrasonics, Ferroelectrics, and Frequency Control}, 2021.

\bibitem{2019Localizing}
R.~J.~V. Sloun and L.~Demi, ``Localizing b-lines in lung ultrasonography by
  weakly supervised deep learning, in-vivo results,'' \emph{IEEE Journal of
  Biomedical and Health Informatics}, vol.~PP, no.~99, 2019.

\bibitem{selvaraju2017grad}
R.~R. Selvaraju, M.~Cogswell, A.~Das, R.~Vedantam, D.~Parikh, and D.~Batra,
  ``Grad-cam: Visual explanations from deep networks via gradient-based
  localization,'' in \emph{Proceedings of the IEEE international conference on
  computer vision}, 2017, pp. 618--626.

\bibitem{kerdegari2021b}
H.~Kerdegari, N.~T.~H. Phung, A.~McBride, L.~Pisani, H.~V. Nguyen, T.~B. Duong,
  R.~Razavi, L.~Thwaites, S.~Yacoub, A.~Gomez \emph{et~al.}, ``B-line detection
  and localization in lung ultrasound videos using spatiotemporal attention,''
  \emph{Applied Sciences}, vol.~11, no.~24, p. 11697, 2021.

\bibitem{2020DL4COVIDinPOC}
R.~Subhankar, M.~Willi, O.~Sebastiaan, L.~Ben, F.~Enrico, S.~Cristiano,
  H.~Iris, C.~Nishith, M.~Federico, S.~Alessandro, P.~Emanuele, T.~Riccardo,
  M.~Giovanni, T.~Elena, I.~Riccardo, S.~Andrea, S.~Gino, R.~Paolo, P.~Andrea,
  v.~{Ruud J. G.}, R.~Elisa, and D.~Libertario, ``Deep learning for
  classification and localization of covid-19 markers in point-of-care lung
  ultrasound,'' \emph{IEEE Transactions on Medical Imaging}, vol.~39, no.~8,
  pp. 2676--2687, 2020.

\bibitem{jaderberg2016spatial}
M.~Jaderberg, K.~Simonyan, A.~Zisserman, and K.~Kavukcuoglu, ``Spatial
  transformer networks,'' 2016.

\bibitem{tsai2021automatic}
C.-H. Tsai, J.~van~der Burgt, D.~Vukovic, N.~Kaur, L.~Demi, D.~Canty, A.~Wang,
  A.~Royse, C.~Royse, K.~Haji \emph{et~al.}, ``Automatic deep learning-based
  pleural effusion classification in lung ultrasound images for respiratory
  pathology diagnosis,'' \emph{Physica Medica}, vol.~83, pp. 38--45, 2021.

\bibitem{2015Detection}
S.~K. Veeramani and E.~Muthusamy, ``Detection of abnormalities in ultrasound
  lung image using multi-level rvm classification,'' \emph{The Journal of
  Maternal-Fetal \& Neonatal Medicine}, 2015.

\bibitem{2020COVID19}
M.~J. Horry, S.~Chakraborty, M.~Paul, A.~Ulhaq, and N.~Shukla, ``Covid-19
  detection through transfer learning using multimodal imaging data,''
  \emph{IEEE Access}, vol.~PP, no.~99, pp. 1--1, 2020.

\bibitem{cheng2021transfer}
D.~Cheng and E.~Y. Lam, ``Transfer learning u-net deep learning for lung
  ultrasound segmentation,'' 2021.

\bibitem{almeida2020lung}
A.~Almeida, A.~Bilbao, L.~Ruby, M.~B. Rominger, D.~L{\'o}pez-De-Ipi{\~n}a,
  J.~Dahl, A.~ElKaffas, and S.~J. Sanabria, ``Lung ultrasound for point-of-care
  covid-19 pneumonia stratification: computer-aided diagnostics in a
  smartphone. first experiences classifying semiology from public datasets,''
  in \emph{2020 IEEE International Ultrasonics Symposium (IUS)}.\hskip 1em plus
  0.5em minus 0.4em\relax IEEE, 2020, pp. 1--4.

\bibitem{born2020pocovidnet}
J.~Born, G.~Brändle, M.~Cossio, M.~Disdier, J.~Goulet, J.~Roulin, and
  N.~Wiedemann, ``Pocovid-net: Automatic detection of covid-19 from a new lung
  ultrasound imaging dataset (pocus),'' 2020.

\bibitem{xue2021modality}
W.~Xue, C.~Cao, J.~Liu, Y.~Duan, H.~Cao, J.~Wang, X.~Tao, Z.~Chen, M.~Wu,
  J.~Zhang \emph{et~al.}, ``Modality alignment contrastive learning for
  severity assessment of covid-19 from lung ultrasound and clinical
  information,'' \emph{Medical Image Analysis}, vol.~69, p. 101975, 2021.

\bibitem{panicker2021approach}
M.~R. Panicker, Y.~T. Chen, K.~V. Narayan, C.~Kesavadas, A.~Vinod
  \emph{et~al.}, ``An approach towards physics informed lung ultrasound image
  scoring neural network for diagnostic assistance in covid-19,'' \emph{arXiv
  preprint arXiv:2106.06980}, 2021.

\bibitem{HEKLER201991}
\BIBentryALTinterwordspacing
A.~Hekler, J.~S. Utikal, A.~H. Enk, W.~Solass, M.~Schmitt, J.~Klode,
  D.~Schadendorf, W.~Sondermann, C.~Franklin, F.~Bestvater, M.~J. Flaig,
  D.~Krahl, C.~{von Kalle}, S.~Fröhling, and T.~J. Brinker, ``Deep learning
  outperformed 11 pathologists in the classification of histopathological
  melanoma images,'' \emph{European Journal of Cancer}, vol. 118, pp. 91--96,
  2019. [Online]. Available:
  \url{https://www.sciencedirect.com/science/article/pii/S0959804919303806}
\BIBentrySTDinterwordspacing

\bibitem{supriya2020machine}
M.~Supriya and A.~Deepa, ``Machine learning approach on healthcare big data: a
  review,'' \emph{Big Data and Information Analytics}, vol.~5, no.~1, pp.
  58--75, 2020.

\bibitem{DLinMUS}
S.~Liu, Y.~Wang, X.~Yang, S.~Li, T.~Wang, B.~Lei, D.~Ni, and L.~Liu, ``Deep
  learning in medical ultrasound analysis: A review,'' \emph{Engineering},
  vol.~5, pp. 261--275, 03 2019.

\bibitem{lecun2021self}
Y.~LeCun and I.~Misra, ``Self-supervised learning: The dark matter of
  intelligence,'' 2021.

\bibitem{jing2020self}
L.~Jing and Y.~Tian, ``Self-supervised visual feature learning with deep neural
  networks: A survey,'' \emph{IEEE transactions on pattern analysis and machine
  intelligence}, vol.~43, no.~11, pp. 4037--4058, 2020.

\bibitem{jamaludin2017self}
A.~Jamaludin, T.~Kadir, and A.~Zisserman, ``Self-supervised learning for spinal
  mris,'' in \emph{Deep Learning in Medical Image Analysis and Multimodal
  Learning for Clinical Decision Support}.\hskip 1em plus 0.5em minus
  0.4em\relax Springer, 2017, pp. 294--302.

\bibitem{mongodi2021quantitative}
S.~Mongodi, D.~De~Luca, A.~Colombo, A.~Stella, E.~Santangelo, F.~Corradi,
  L.~Gargani, S.~Rovida, G.~Volpicelli, B.~Bouhemad \emph{et~al.},
  ``Quantitative lung ultrasound: technical aspects and clinical
  applications,'' \emph{Anesthesiology}, vol. 134, no.~6, pp. 949--965, 2021.

\end{thebibliography}

\end{document}